\documentclass[aps,amsmath,amssymb,reprint]{revtex4-1}
\usepackage{graphicx}
\usepackage{dcolumn}
\usepackage{bm}
\usepackage[utf8]{inputenc}
\usepackage{natbib}
\usepackage[T1]{fontenc}
\usepackage{mathptmx}
\usepackage{etoolbox}

\makeatletter 
\def\@email#1#2{%
 \endgroup
 \patchcmd{\titleblock@produce}
  {\frontmatter@RRAPformat}
  {\frontmatter@RRAPformat{\produce@RRAP{*#1\href{mailto:#2}{#2}}}\frontmatter@RRAPformat}
  {}{}
}%

\usepackage{hyperref}
\hypersetup{colorlinks,linkcolor={blue},citecolor={blue},urlcolor={blue}}
\begin{document}
\preprint{AIP/123-QED}

\title {Global optimisation of the control strategy of a Brownian information Engine: Efficient information-energy exchange in a generalised potential energy surface}

\author{Rafna Rafeek}
\affiliation{Department of Chemistry, Indian Institute of Technology Tirupati, Tirupati, Andhra Pradesh 517619, India}
\author{Debasish Mondal}
  \email{rafna.rk@gmail.com, debasish@iittp.ac.in}
\affiliation{Department of Chemistry, Indian Institute of Technology Tirupati, Tirupati, Andhra Pradesh 517619, India}

\date{\today}
\begin{abstract}
An information engine harnesses energy from a single heat bath, utilising the gathered information. This study explores the best control strategy of a Brownian information engine (BIE), confined in a potential energy surface (PES) of arbitrary shape, and experiencing a measurement outcome-based feedback cycle. The feedback site corresponds to an instantaneous shift in the potential centre to an additional feedback distance over the measurement outcome. The strategy for the most efficient information-to-energy conversion is achieved when the position of the global potential minimum corresponds to the additional feedback distance. The BIE acts as a heater if and only if the average potential energy is higher than the energy at the additional feedback distance. Operating under confinement PES of different shapes, the BIE can harness energy beyond the average potential energy, and multiple heater-refrigerator re-entrance events are feasible. The consequences of the best control strategy are explained using sufficient examples.
\end{abstract}
\maketitle

\section{INTRODUCTION}
Information-energy exchange in fluctuating environments has garnered significant interest in recent years due to its relevance in uncovering the operating principles behind biological motors \cite{Heiner_2021, borsley_2021, chakraborty2023, barato2013, barato2014, Horowitz2013}.  Various cellular processes that occur on a mesoscopic scale and in the presence of inherent thermal and active fluctuations \cite{fang2019, xiao2009, Seifert_2012} operate in the presence of a single heat reservoir. Using the principles of thermodynamics, the understanding of the underlying physics of such processes is thus not so straightforward. Information thermodynamics bridges this gap and addresses the possibility of energy harnessing from a mesoscopic-scale thermodynamic system coupled to a single thermal bath. The issue traces back to the idea of Maxwell's demon \cite{maxwell1871} and Szilard's engine \cite{Szilard1929zphys}. The archetypal Maxwell demon-like feedback controller can use acquired information to extract work \cite{maxwell1871}. The operational sequence of information engines typically involves a feedback process based on the acquired information, wherein the system response enables the work outputs \cite{  rafeek2024, archambault2024,  Paneru_2022, Park2016pre, Paneru2020natcommun, Paneru2018prl, Abreu2011epl, Abreu2012prl, Pal2014pre, paneru2018pre, rafeek2023geometric, rafna_pot_2_2025, ali2022geometric, rafna_pot_2025,
Ashida2014pre, Bauer_2012, Mandal_2013, malgaretti2022, Taichi_2013, Kim2011prl, Goold2016jphysA, koski2014pnas, Toyabe2010natphys}. Furthermore, improved theoretical understanding and experimental control of mesoscale stochastic processes have paved the way for studies of Brownian information engines (BIE) \cite{Abreu2011epl, Abreu2012prl, Bauer_2012, Pal2014pre, Ashida2014pre, Park2016pre, archambault2024, Paneru2020natcommun, Paneru2018prl, Paneru_2022, paneru2018pre, ali2022geometric, rafeek2023geometric, rafna_pot_2025, rafna_pot_2_2025} for the last one and a half decades. 

BIEs are often modeled with sets of conditions that consist of Brownian particles confined in a harmonic potential (in 1-D), immersed in a thermal bath, and subjected to feedback-driven cycles \cite{Abreu2011epl, Abreu2012prl, Bauer_2012, Pal2014pre, Ashida2014pre, Park2016pre, Paneru2020natcommun, Paneru2018prl, Paneru_2022, paneru2018pre, rafna_pot_2_2025, rafna_pot_2025,ali2022geometric, rafeek2023geometric}. The feedback is implemented by shifting the centre of the harmonic potential based on the positional outcome, thereby enabling work extraction. In general, the shape of the confinement \cite{ali2022geometric, rafeek2023geometric, rafna_pot_2025, rafna_pot_2_2025}, extent and nature of the environmental fluctuations \cite{Paneru_2022, rafeek2024, malgaretti2022}, and the chosen feedback strategy \cite{Abreu2011epl, Ashida2014pre, Taichi_2013} are the three crucial factors that influence the performance of a BIE.
Because of their academic utility and/or easy experimental implementation, confining potentials are preferably chosen to be a harmonic one and two distinct types of feedback strategies are often used. We name them as: symmetric, allowing a symmetric bidirectional shift of the potential centre \cite{Ashida2014pre, ali2022geometric, rafna_pot_2025}. In this case, the particle position is measured, and the centre of the potential is shifted to the measurement location instantaneously, then allowed to be relaxed.  On the other hand, an asymmetric feedback is unidirectional \cite{Abreu2011epl, Abreu2012prl, paneru2018pre, Pal2014pre, Paneru2018prl, rafeek2023geometric, rafeek2024, Paneru_2022, archambault2024,rafna_pot_2_2025, Paneru2020natcommun, paneru2018pre, Park2016pre}. A measurement distance is first set by the observer. If the particle is beyond that measurement distance, the confinement centre is shifted to a targeted feedback location. A symmetric feedback controller can estimate the maximum achievable energy output for the given feedback cycle ($\frac{1}{2} k_BT$ for a BIE working in a harmonic trap) by accurately calculating the net information gain \cite{Ashida2014pre}. In contrast, an asymmetric protocol is easy to implement in the experiment, and engine performance is strongly influenced by the spread of the equilibrium distribution, which in turn depends on the shape of the confinement and noise strength\cite{Park2016pre, Paneru_2022, Paneru2018prl}.
    
 Our recent work \cite{rafeek2024} and \cite{Paneru_2022} demonstrate the colossal performance of the BIE in the presence of an active bath in addition to a thermal reservoir.  Moreover, we have examined how the shape of potential confinement affects the engine's operational requisites for both the commonly used feedback protocol \cite{rafna_pot_2025, rafna_pot_2_2025}. However, both of the feedback controllers follow a predefined, distinct set of sequential steps. 
\begin{figure}[!h]
\centering
    \includegraphics[width=0.5\textwidth]{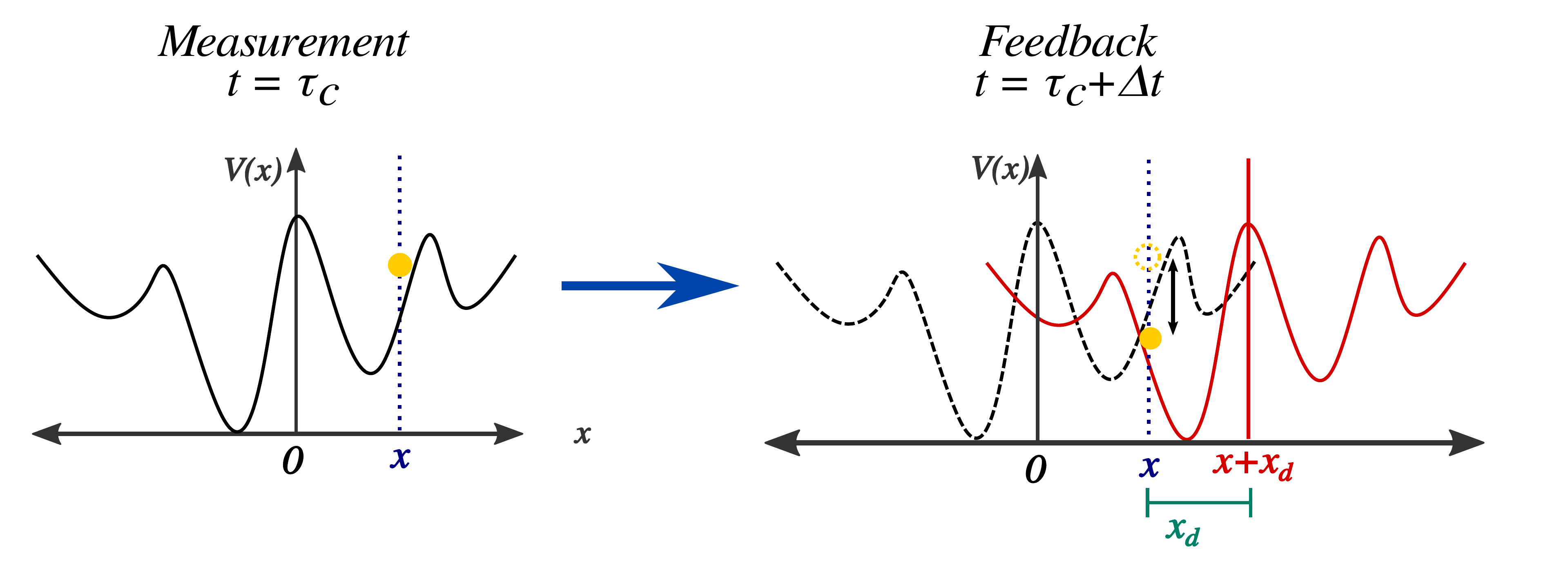} 
     \caption{Schematic representation of the feedback strategy with a Brownian particle trapped in an arbitrary potential. In the measurement step, particle position is obtained $(x)$ accurately. In the feedback step, the potential centre is instantly shifted to a new feedback site, $x_f=x+x_d$.
     }
    \label{f1}
\end{figure}
For both types of commonly used feedback strategies, the practical interest lies in maximising the output work from a single thermal (fluctuating) bath. During the feedback step, the centre of the confining potential is shifted to the targeted feedback site. Such feedback results in a loss of the surprise (gain in information) of the particle's finding probability if the potential centre is itself an energy minimum (which is the case for a harmonic potential). Therefore, the parameters of the feedback controllers are conveniently adjusted to reduce the system energy during sudden feedback, thereby enabling work extraction. Beyond a specific feedback site, the cycle may experience a high information loss during relaxation, resulting in refrigeration.

Now, instead of being a harmonic trap, if the confining potential surface of the BIE has an arbitrary shape, with multiple maxima and minima of different energies, a smart feedback strategy is required to achieve the best engine efficiency and to identify the phase space that the heater-refrigerator. Therefore, tuning the protocol itself to find the universal feedback strategy that maximizes the efficiency of the information-energy exchange of a BIE that operates in an arbitrary confining potential remains unexplored and warrants considerable attention. It is worth mentioning that the stochastic movement of Brownian particles in a bounded potential surface with multiple hills and valleys is of significant interest in understanding many biological processes and set-up, such as gene regulatory networks \cite{ozbudak2004}, cell signaling pathways \cite{angeli2004}, metabolic networks \cite{craciun2006}, and even in ecosystems \cite{bastiaansen2018}. Moreover, nonequilibrium phenomena like noise-assisted barrier crossing rate \cite{marie2020, mondal2011asymmetric}, and different noise-induced phenomena like stochastic resonance  \cite{mcnamara1989,burada2008entropic, mondal2016resonance, ghosh2007, zhu2022, ali2024}, resonant activation \cite{doering1992, reimann1995, jan1996, wagner1999, bart2002, mondal2010activation, mondal2010resonant, fisconaro2011, chattoraj2014dynamics, mondal2012entropic, ray2014} and ratchet rectification \cite{magnasco1993, zwanzig1988, mondal2009roughratchet, kato2013quantum, mondal2016ratchet} observed in varied potential landscapes, reveals non-trivial insights into the essentials of the underlying physics. Motivated by these considerations, we investigate the generalised optimal feedback strategy for best information-to-energy conversion, independent of the nature of the confinement. In addition, we also aim to understand the operational feedback mechanism that leads to heating-refrigeration crossover and sometimes their re-entrances, when the BIE is operating in such an arbitrary potential landscape.  

\section{Finding the optimal feedback strategy}  
Let us consider non-interacting overdamped Brownian particles in 1-D, confined in an external potential, $V(x-\lambda$), centred at $\lambda$ that follows the dimensionless Langevin equation of motion: 
\begin{equation}\label{1}
    \begin{aligned}
          \gamma \dot{x} &= -V'(x-\lambda) + \sqrt{2D} \gamma \zeta(t), \\
         \text{with} \;\; \left \langle \zeta (t)\right \rangle &= 0, \;\; \left \langle \zeta(t) \zeta ({t}')\right \rangle =\delta (t-{t}'), 
    \end{aligned}
\end{equation}
where $\gamma$ denotes friction coefficient (taken as unity), $D$ is the diffusion coefficient $(D=k_BT / \gamma)$, $T$ refers to the temperature in Kelvin scale, and $k_B$ represents the Boltzmann constant. Unless mentioned otherwise, we consider $k_BT=1$. Thermal fluctuations are mapped by zero-mean Gaussian white noise $\zeta(t)$. The stationary state equilibrium distribution of particle position $(P_{eq}(x))$ follows the Boltzmann distribution:
\begin{equation}\label{2}
    \begin{aligned}
        P_{eq}(x) = \mathcal{N} \exp \bigg [ -\frac{V(x -\lambda ) }{k_B T} \bigg],
    \end{aligned}
\end{equation}
where $\mathcal{N}$ is the normalization constant. We consider a bounded potential which practically confines the particle within a finite domain (of $x$), i.e. $P_{eq}(x) \to 0, \text{ if } x \to \pm \infty$.
Further, consider a general feedback controller that consists of three steps: measurement, feedback and relaxation. The feedback is employed to the thermally equilibrated system, which is confined within an arbitrary potential $(V(x))$ centred at the origin $\lambda=0$. In the measurement step $(\text{at\;} t = \tau_c)$, we gather the positional information (accurate) of the particle $(x)$. In the feedback step $(\text{at\;} t = \tau_c + \Delta t)$, the potential centre is promptly shifted to a new feedback location $(x_f)$. Here we consider the feedback location as the sum of the positional outcome $(x)$ and a predefined additional feedback distance $(x_d)$, i.e. $x_f =x + x_d$ (schematically shown in Fig.~\ref{f1}). Finally, we let the particle relax in the shifted potential $(\text{during\;} \tau_c + \Delta t < t < 2 \tau_c)$. The potential shift is faster than the thermal relaxation time $(\tau_r)$, such that there is no heat dissipation. Consequently, the sudden change in the potential energy can completely be converted into work extraction: $-W(x) = V(x)-V \left( x-x_f \right)$. Therefore, the work extracted from the single feedback cycle takes the form:  $-W(x)= V(x)-V(-x_d)$.
The cycle is repeated to find the average extractable work:
\begin{equation}\label{4}
      -\langle W \rangle = -\int_{-\infty}^{\infty} dxP_{eq}(x)W(x)
       = \langle V(x) \rangle - V(-x_d).
\end{equation}
  
 Thus, the efficiency of the feedback mechanism is characterised by the average potential energy of the confined particle and the additional feedback distance, $x_d \in [-\infty, \infty]$. For simplicity, we focus our discussion on the centrosymmetric potential expressed by even functions only, which restricts $x_d \in [0, \infty]$ and is sufficient for the study and ensures $V(-x_d)=V(x_d)$. We assure the fact that all the principal outcomes of this study are equally true for an asymmetric potential surface as well.
 
 We recall that the main goal of the paper is to establish the best feedback strategy in terms of efficient information harnessing by a BIE operating in a potential of arbitrary shape. The average potential energy $(\langle V(x) \rangle)$ of the system remains constant for a given potential. So, the work extracted from a BIE can be modulated by varying the additional feedback distance $(x_d)$.
Clearly, with a decrease in the potential energy at the additional feedback distance $(V(x_d))$, the amount of work extracted through the feedback increases $(-\langle W \rangle)$. For a bounded confinement, $x_{min}$ corresponds to the position coordinate of the potential stablest point. Thus, maximum work can be extracted from the feedback mechanism with $x_d=x_{min}$, for potential of any nature.
Therefore, to achieve maximum work extraction, the generalised optimal strategy involves shifting the potential centre to the feedback location $x_f^*=x+x_{min}$. Thus, the maximum work extracted $(-\langle W \rangle^{max})$ depends on the stability of the confinement well(s) ($V(x_d$) and the average potential energy $(\langle V(x) \rangle)$ under consideration. Consider a situation with a positive average potential energy
$\langle V(x) \rangle >0$. For a feedback control, where the potential at the additional feedback distance is less than the average potential energy $(V(x_d) < \langle V(x) \rangle)$, the BIE functions as an engine. Refrigeration is observed otherwise. The heater-to-refrigeration (or vice versa) transition occurs at the inverse additional distance $x_d^{inv}$, which corresponds to no work output $(-\langle W \rangle=0)$. Interestingly, depending on the spatially varying confinement profile, multiple feedback settings can result in zero work extraction. Now, as we have considered a bounded potential, i.e. $V(x)(x) \to +\infty, \text{ if } x \to \pm \infty$, the situation $\langle V(x) \rangle  < 0$ arises only when $V(x_d=x_{min}) < \langle V(x) \rangle$. Thus, the essential criterion for a BIE to be an engine is always $V(x_d=x_{min}) < \langle V(x) \rangle$. This also ensures that when the distance between the potential centre and the global minimum ($x_{min}^G$) of the PES is equal to the additional feedback distance ($x_d=x_{min}^G$), the exchange of information-energy is the maximum.

The information gathered during the measurement is equal to the uncertainty or surprise in the outcome of the event. The event with low probable outcomes tends to have higher acquired information than one with high probable outcomes. 
Following the standard definition \cite{parrondo2015}, the average information acquired during the measurement process related to the position $x$ is:
\begin{equation}\label{5}
    \begin{aligned}
       \langle I \rangle = -\int_{-\infty}^{\infty} dxP_{eq}(x)\ln[P_{eq}(x)] = - \ln \left[ \mathcal{N} \right] + \frac{\left\langle V(x) \right\rangle}{k_BT}.
    \end{aligned}
\end{equation}
However, some of the acquired information will be lost during the thermal relaxation. The average unavailable information \cite{Ashida2014pre} can be estimated as:
\begin{equation}\label{6}
    \begin{aligned}
       \langle I_u \rangle &= -\int_{-\infty}^{\infty} dxP_{eq}(x)\ln[P_{eq}(x-x_f)] \\ &= -\int_{-\infty}^{\infty} dxP_{eq}(x)\ln[P_{eq}(x_d)] = - \ln \left[ \mathcal{N} \right] + \frac{ V(x_d) }{k_BT}.
    \end{aligned}
\end{equation}
Compared with Eqs.~\ref{4}-\ref{6}, the present protocol achieves complete conversion of the available information into work extraction: $-\langle \overline {W} \rangle =  \langle I \rangle -\langle I_u \rangle$. 
One could also confirm that the feedback protocol satisfies the integral fluctuation theorem by plugging in the corresponding expressions for work and information as:
\begin{equation}\label{7}
\begin{aligned}
      \langle \text{e}^{-(\overline{W}+I-I_u )} \rangle =
    \int_{-\infty}^{\infty} {dx} P_{eq}({x})\text{e}^{\overline{V}(x)-\overline{V}(x_d)}\frac{P_{eq}({x})}{P_{eq}(x_d)}
     \;\;= 1.
\end{aligned}
\end{equation}
Where $\overline{Z}$ represents the scaled parameter with respect to thermal energy, i.e. $\overline{Z} = \frac{Z}{k_BT}$.
The definitions imply that for a given confining potential, the acquired information $(\langle I \rangle)$ is determined solely by the dispersion of the equilibrium distribution. However, the details of the feedback strategy influence the amount of information lost $(\langle I_u \rangle)$ during relaxation. Thus, it is straightforward that the feedback strategy with the least information loss will exhibit maximum information-energy conversion in a BIE. The information lost during the relaxation depends on the potential energy value at the additional feedback distance $(V(x_d))$. Any feedback with a highly stable $(V(x_d))$ (generally a high negative) reduces the loss of information during the relaxation step. 
The information lost is minimum for the choice of $x_d$ at the global potential minimum, i.e., $x_f^{*}=x_{min}^G$ or $x_d=x_{min}^G-x$. Therefore, the best strategy to achieve the upper limit on information-energy exchange is to set the feedback site at $x_f^* = x+x_{min}^G$, and a condition $(\langle I \rangle > \langle I_u \rangle)$ ensures that the feedback controller acts as an engine.
 In a similar lane, when feedback involves highly unstable energy at the additional feedback distance, the resulting information loss may exceed the acquired information $(\langle I \rangle < \langle I_u \rangle)$, causing the engine to operate as a refrigerator.  
To summarize the main results, the optimal strategy for achieving the best information-energy exchange dictates the choice of the feedback site at $x_f^*=x+x_{min}^G$, where $x_{min}^G$ denotes the position of the global potential minima. We find that the upper bound of work extraction $(-\langle W \rangle^{max})$ can be tuned with the nature of the confinement. Moreover, certain potential profiles can yield multiple conditions for heater-to-refrigerator transitions, indicating re-entrance behaviour. In the recent study \cite{rafna_pot_2025, rafna_pot_2_2025}, we have shown the influence of concavity and varying stability of confinement on information-energy exchange in a fluctuating system. In the following, we employ the best feedback strategy in looking into its consequences with three different centrosymmetric potential surfaces: (a) a monostable trap with varying concavity, (b) a bistable potential with unstable potential centre, and (c) a triple-well energy surface with modulating multistability, as illustrative examples.  
\begin{figure}[!htb]
\centering \includegraphics[width=0.42\textwidth]{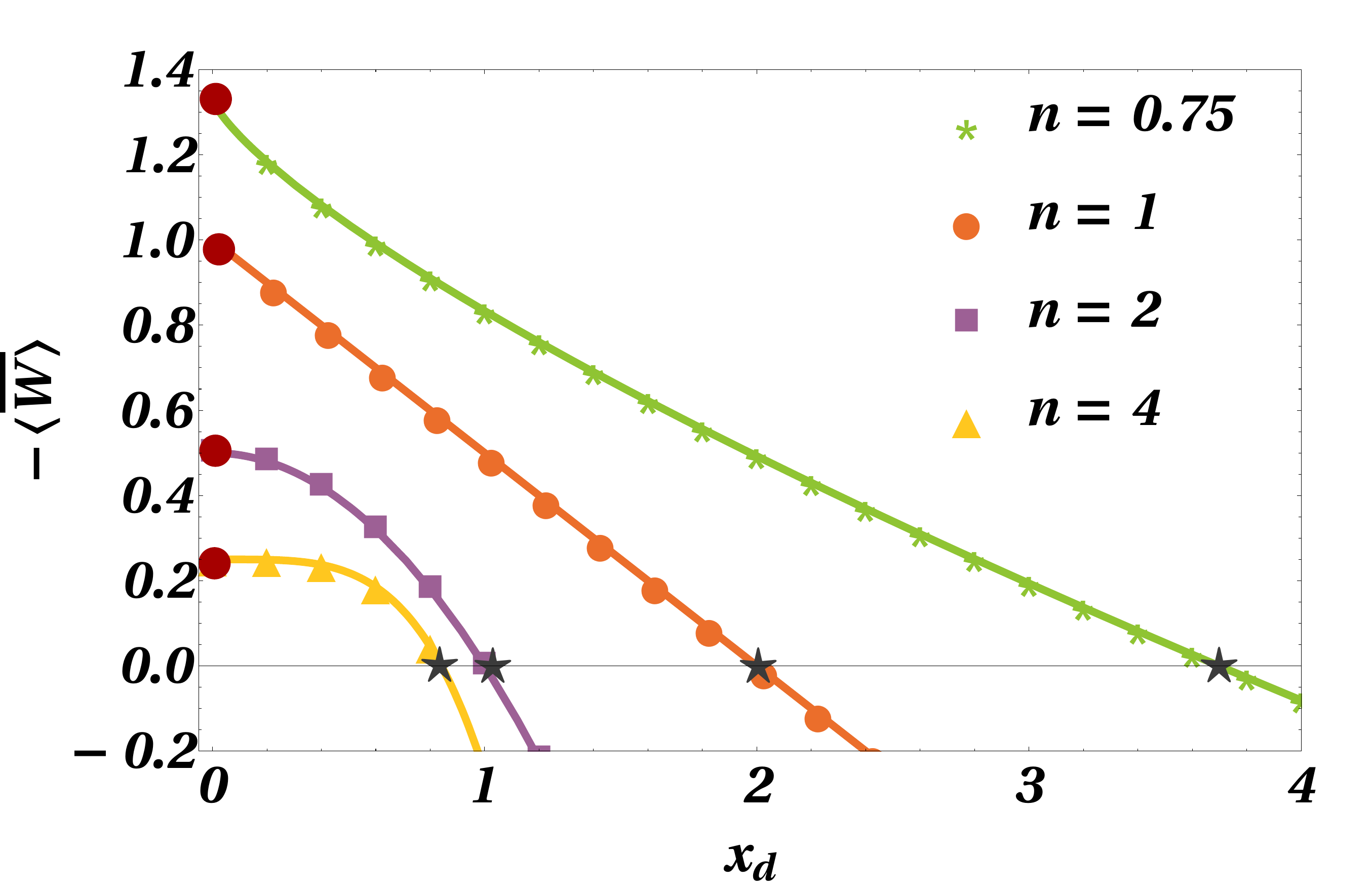}
   \caption{Variation of scaled extractable work $(-\langle \overline{W} \rangle)$ as a function of additional feedback  distance $(x_d)$, for different values of $n$ and for $a = 0.5$. The red-colored filled circle indicates the best feedback locations $(x_f^*)$ that correspond to $-\langle W \rangle^{max}$, and the dark grey colored filled star shows the heater-refrigerator transition point ($x_d^{inv}$).}
    \label{f2}
\end{figure}

\section{Examples to explain Information-Energy conversion efficiency}
\subsection{BIE with a monostable trapping of different concavity: Optimal strategy and the upper bound of output}

To begin with, we consider a monostable potential energy centred at the origin, and expressed as the power law $ V(x) = a|x|^n$, where $a$ and $n$ are the force constant and the exponent, respectively (both $a,\;n > 0$). The shape of the potential with different extents of concavity is shown in Fig.~\ref{fa1}(a). The equilibrium distribution function $P_{eq}(x)$ follows the symmetric unimodal distribution, but overall dispersion gets wider with increasing concavity ($1 < n$ and decreasing), as shown in Fig.~\ref{fa2}.(a).
Using Eq.~\ref{4}, the achievable work output from the engine operating in such a monostable confinement can be obtained as
    $    - \langle \overline {W} \rangle = \frac{1}{n} - \overline{a} |x_d|^n $.

Therefore, the maximum work extraction is possible for feedback with $x_d=0$, i.e. $ -\langle \overline {W} \rangle ^{max}$ at $x_f^* = x$. Any nonzero values of the additional feedback distance $(x_d \neq 0)$ reduce the average work output because of the rise in potential energy at the additional distance $V(x_d) >0$. Fig.~\ref{f2} shows a monotonic decrease in average work output as a function of $x_d$ for confinement with different concavities. Clearly, the variation of achievable work with a negative $x_d$ has a mirror image symmetry to the positive side.
Thus, the result rightly reproduces the criteria optimal feedback strategy used in a symmetric feedback controller for achieving the upper bound of output in a monostable potential trapping \cite{Ashida2014pre,rafna_pot_2025}. The results also depict that the upper limit of work extraction equals the average potential energy \cite{Ashida2014pre,rafna_pot_2025} or the average of effective entropic potential \cite{zwanzig1992, reguera2001, reguera2006entropic, mondal2010entropic, das2012shape, mondal2011, das2012logic, ali2022geometric} of the confining setup $\frac{1}{n}$. We also find that the protocol criteria for heater-to-refrigeration transition ($\overline {W} \rangle=0$), is  achieved at $x_d^{inv}=(an)^{-\frac{1}{n}}$. Consequently, the engine operates as a refrigerator for $x_d > (an)^{-\frac{1}{n}}$. The transition points are shown using filled stars in Fig.~\ref{f2}.

Further, to shed light on the influence of the feedback strategy on the available information, we analyze the variation of average acquired information $(\langle I \rangle)$ and unavailable information $(\langle I_u \rangle)$ as a function of $x_d$. We find that the measured $\langle I \rangle$ and $\langle I \rangle_u$ hold equality in the information-energy relation, $- \langle \overline {W} \rangle = \langle I \rangle - \langle I_u \rangle$ for any arbitrary value of $x_d$. As per definition, $\langle I \rangle$ is expected to be independent of the feedback strategy, while $\langle I_u \rangle$ showcases non-monotonic dependence on the choice of $x_d$ (Appendix Fig.~\ref{fb1}(a)). The minimum information lost is obtained for the optimal feedback condition, i.e., $x_d=0$. Interestingly, under optimal protocol ($x_d=0$) and with an increase in concavity of the confinement (lower $n$), both $\langle I \rangle$  and $\langle I_u \rangle$ increase individually, so does their difference ($\langle I \rangle - \langle I_u \rangle$). 
This can be explained in terms of the increasing dispersion of the particle's position with rising concavity and the relative stability at the potential minimum  $x=x_{min}$, respectively. Finally, one can notice that the feedback condition for $ \langle I \rangle = \langle I_u \rangle $ varies with the concavity of trappping as $x_d^{inv}=(an)^{-\frac{1}{n}}$, yielding the heater-to-refrigeration transition beyond this additional feedback distance. 

\subsection{A perturbed potential centre: Exploring the Heater-Refrigerator phase space}

\begin{figure}[!htb]
    \includegraphics[width=0.4\textwidth]{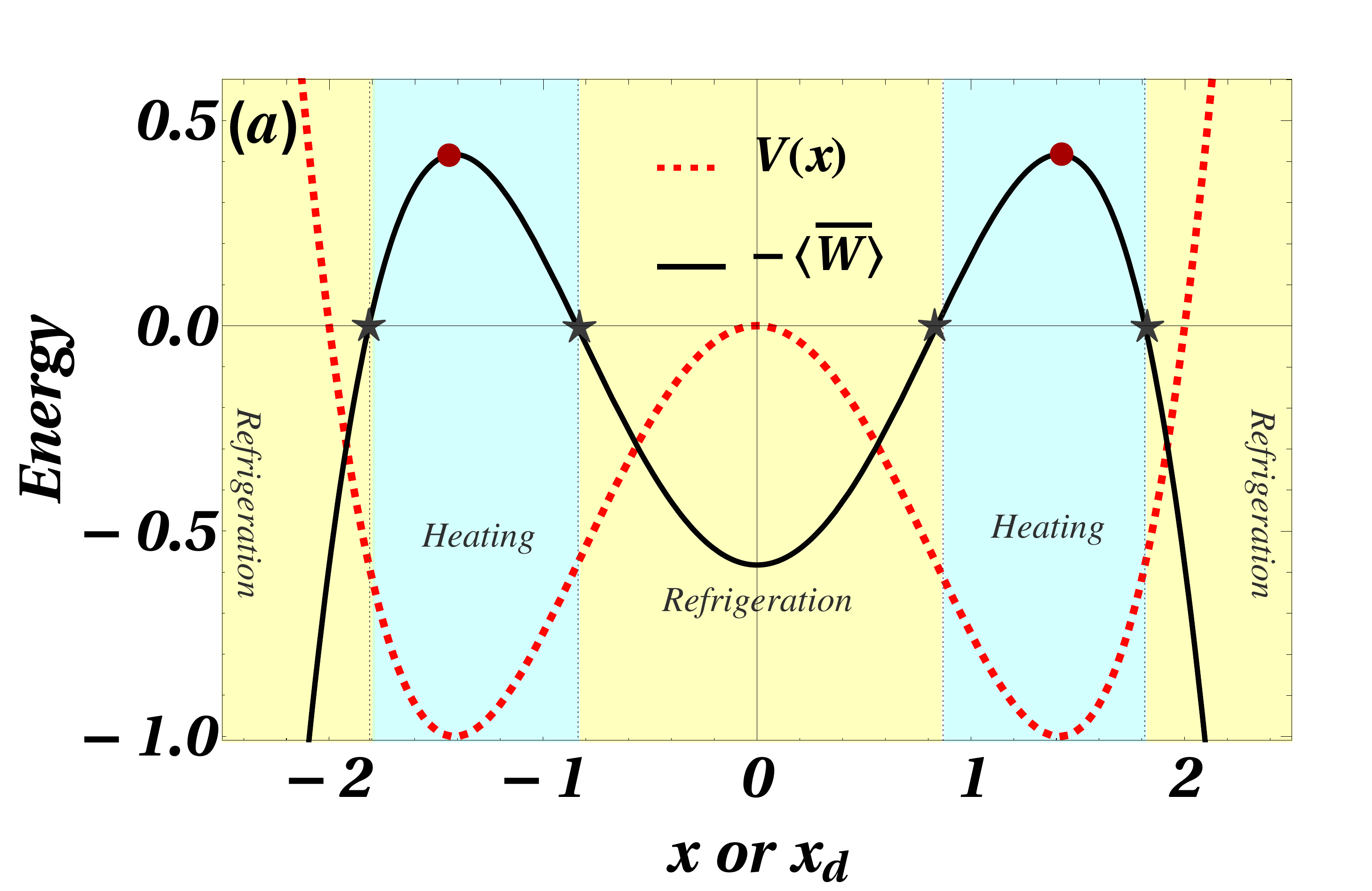} \\
      \includegraphics[width=0.4\textwidth]{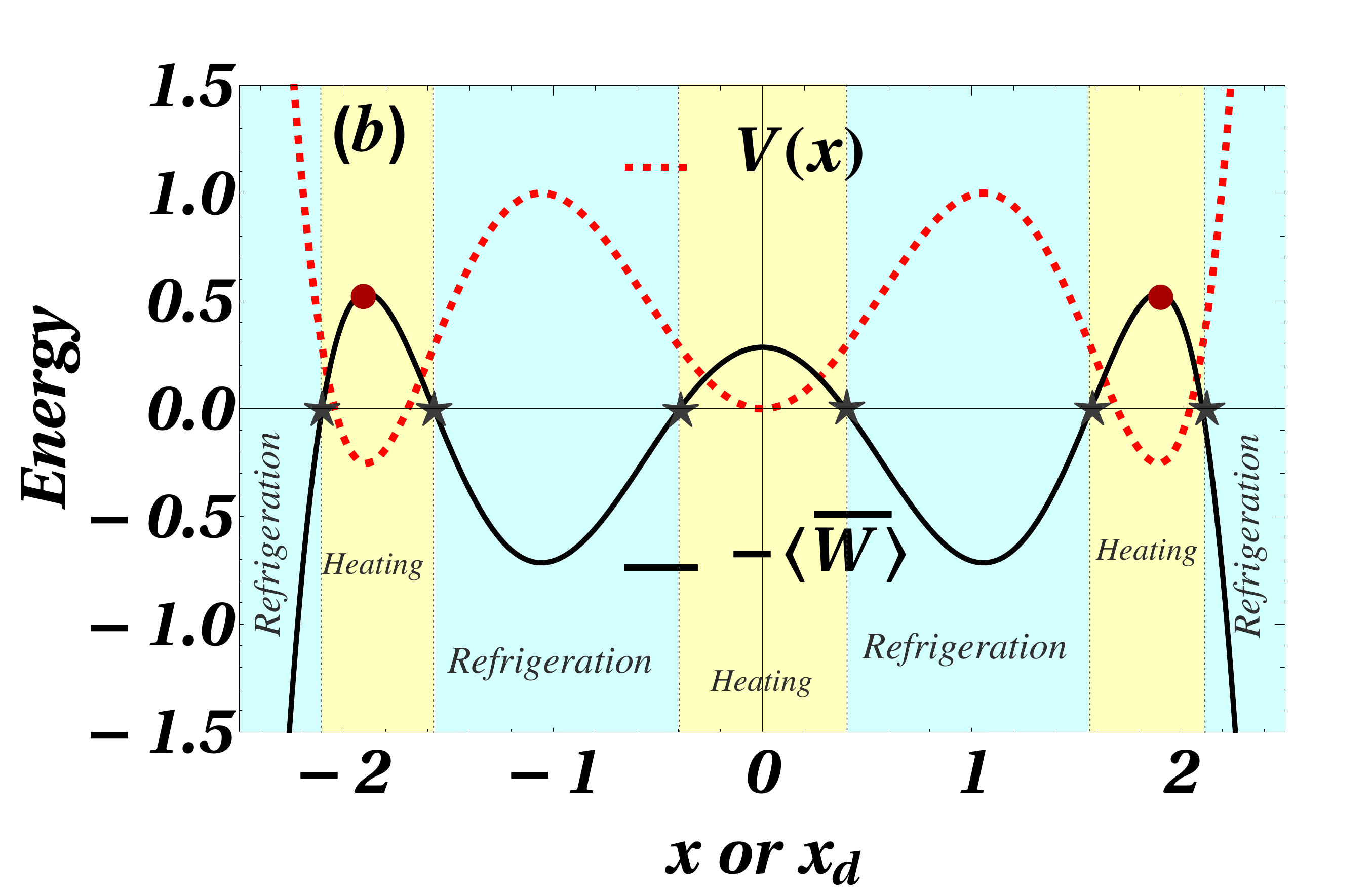} \\
  \caption{A phase diagram of output work showing heater-to-refrigerator transition along with the potential energy $(V(x))$ for: (a) double-well potential with $b=1$ and $\Delta \overline{E} = 1$. (b) triple-well potential with $a=4$, $b=5$ and $c=1$. For both cases, the red-solid circle corresponds to $-\langle W \rangle^{max}$ at $\pm x_{min}$. The grey-solid star corresponds to heater-to-refrigeration transitions at $x_d^{inv}$.}
   \label{f4}
\end{figure}

Next, we consider an instability at the potential centre, and choose a BIE in a bistable potential: $V(x)= -\frac{a}{2}x^2 + \frac{b}{4} x^4$, where $a$ and $b$ are constant (positive) confinement parameters. With $a>0$, the confinement consists of two symmetric wells and a top potential barrier in the centre ($x_{max}=0$) with an energy barrier $\Delta E = a^2/4b$ (Fig.~\ref{fa1}(b)). Two minimum are equally stable and located at a distance $x_{min}=\pm \sqrt{\frac{a}{b}}$.  An increase in the energy barrier elevates the perturbation at the centre, shifts the symmetric wells away from each other, which causes a bimodal distribution with higher dispersion (Fig.~\ref{fa2} (b)).
In the limit of $\Delta\overline  E\rightarrow 0$, the particle becomes localised near the potential centre, as the surface behaves like a rough monostable potential. In contrast, with $\Delta \overline {E} \to \text{ very \;high}$, the confinement acts as two harmonic traps centred near $\pm x_{min}$ (Fig.~\ref{fa2} (b)). 
\begin{figure}[!htb]
    \includegraphics[width=0.4\textwidth]{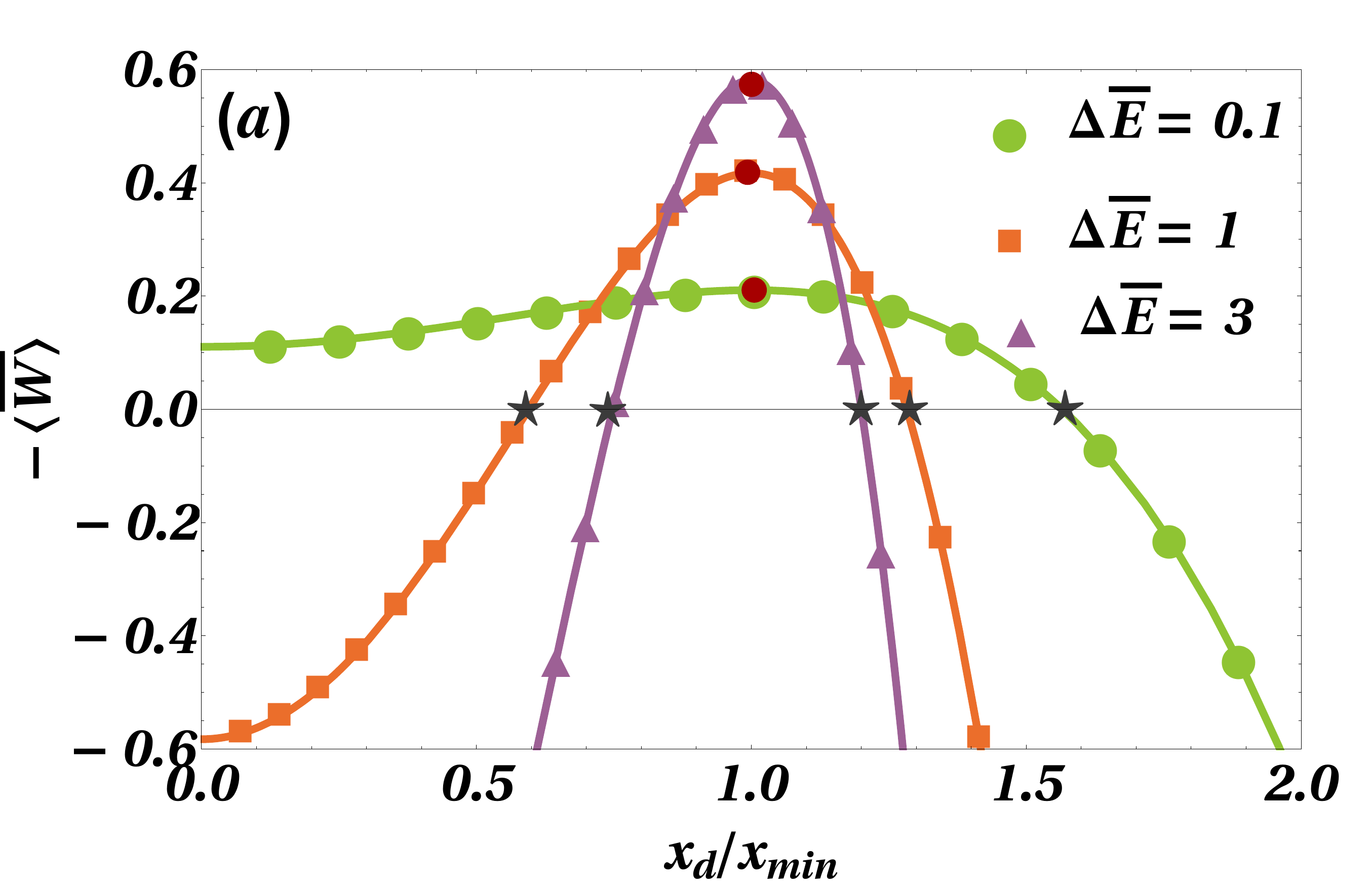} \\
     \includegraphics[width=0.4\textwidth]{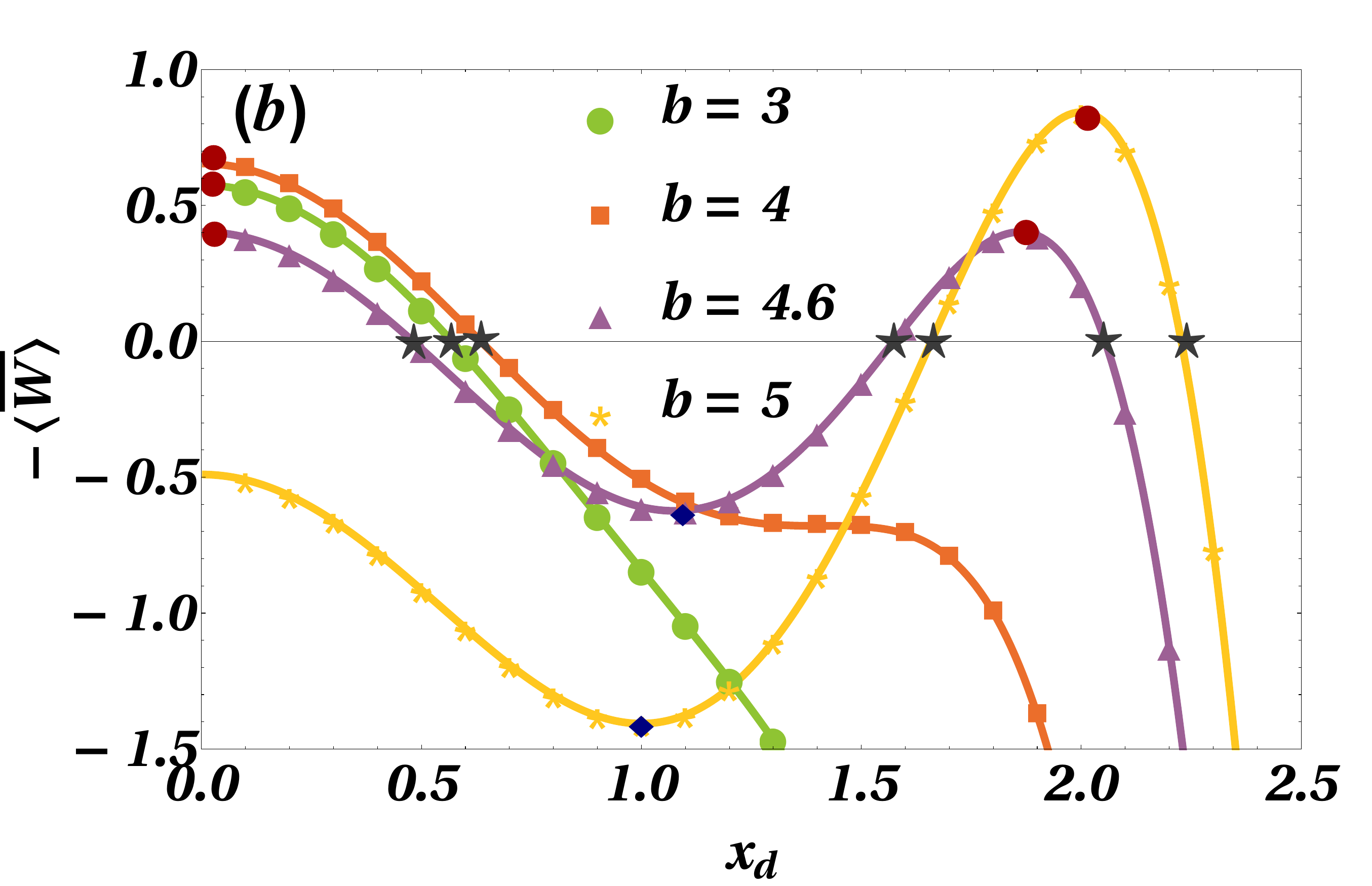} \\
     \caption{ Variation of a scaled output work $(\langle \overline{W} \rangle)$ as a function of  (a) scaled additional distance $(x_d/x_{min})$ for bistable potential with varying $\Delta \overline{E}$ and with $b=1$; (b) additional feedback distance $(x_d)$ for tri-stable potential with varying quartic contribution $(b)$. The parameter set is chosen: $a=4$, $c=1$.
     Red-colored circles indicate additional distance $(x_d^*)$ criteria to obtain work maxima $(-\langle \overline{W} \rangle^{max})$ and dark grey-colored stars represent the output inversion points $(x_d^{inv})$. }
    \label{f3}
\end{figure}
Using Eq.~\ref{4}, and in the presence of such a bistable potential, one can find:
\begin{equation}\label{9}
\begin{aligned}
\langle \overline{ V } (x) \rangle =
& \frac{[4\epsilon - 1] I_{-\frac{1}{4}} \left( \epsilon \right)
-[4 \epsilon +1] I_{\frac{1}{4}}\left(\epsilon\right)}
{4  [ I_{\frac{1}{4}}\left( \epsilon \right ) +I_{-\frac{1}{4}}\left( \epsilon \right ) ]} -\frac{ \epsilon [ I_{\frac{3}{4}}( \epsilon )+I_{\frac{5}{4}}\ ( \epsilon ) ]}{I_{\frac{1}{4}}( \epsilon )+I_{-\frac{1}{4}} (\epsilon)},\\
\overline{V} (x_d) &=    \frac{\overline{b}}{4} x_{d}^4 - \sqrt{2\overline{b}\epsilon} x_d^2,
 \end{aligned}
\end{equation}
where $I_{\nu}(z)$ is the modified Bessel function of the first kind. Therefore, $-\langle W \rangle$ $(=\langle V(x) \rangle - V(x_d))$ can in principle be positive or negative depending on the choice of the additional feedback distance, which leads to heating or refrigeration, respectively. Fig.~\ref{f3}(a) depicts a non-monotonic variation of work extracted $(-\langle \overline{W} \rangle)$ with increasing additional feedback distance, which is scaled as $x_d / x_{min}$, for different extents of centre perturbation (variations with positive $x_d$ are shown here). In this case, $x_{min}^G=\pm x_{min}$ irrespective of the magnitude of the $\Delta E$. Thus, as expected, the maximum work output $(-\langle W \rangle^{max})$ is obtained for a feedback:  $x_d=\pm x_{min}$, or $ x_f^* = x \pm x_{min}$.  We also find that BIE with pronounced instability at the centre may exhibit more than one set of feedback criteria for refrigeration. To explore this further, we plot the phase diagram of the output work within the entire range of feedback site ($-\infty< x_d<\infty$) for an arbitrarily chosen bistable potential (Fig.~\ref{f4}(a)). The phase diagram shows that a BIE may showcase more than one set of $x_d^{inv}$, leading to the refrigerator-to-heater reentrance phenomenon. With a pronounced barrier height ($\Delta\overline {E} \gg 1$), the engine functions as a heater, either of the choices of $x_d$ in the vicinity of the position of the basins $(x_{min})$. It acts as a refrigerator otherwise.  Feedback mechanisms positioned either closer to the perturbed potential centre $(x_d \to 0)$, or far away from the stable centre  $(x_d \gg x_{min})$ function as refrigeration processes. Interestingly, with this high barrier height, the conventional symmetric protocol serves as the criterion for a maximum cooling scenario. A careful scrutiny of the related information changes reveals that the information lost ($I_u$) is minimized for the optimal protocol with feedback location $x_f^*=x+x_{min}$ (Fig.~\ref{fb1} (b)). Also, the confinements with deeper wells portray multiple sets of feedback criteria $(x_d^{inv})$, where $\langle I_u \rangle$ and $\langle I \rangle$ cross each other (Fig.~\ref{fb1} (b)), yielding refrigeration-to-heating re-entrance phenomena. \\

\begin{figure}[!htb]
    \includegraphics[width=0.4\textwidth]{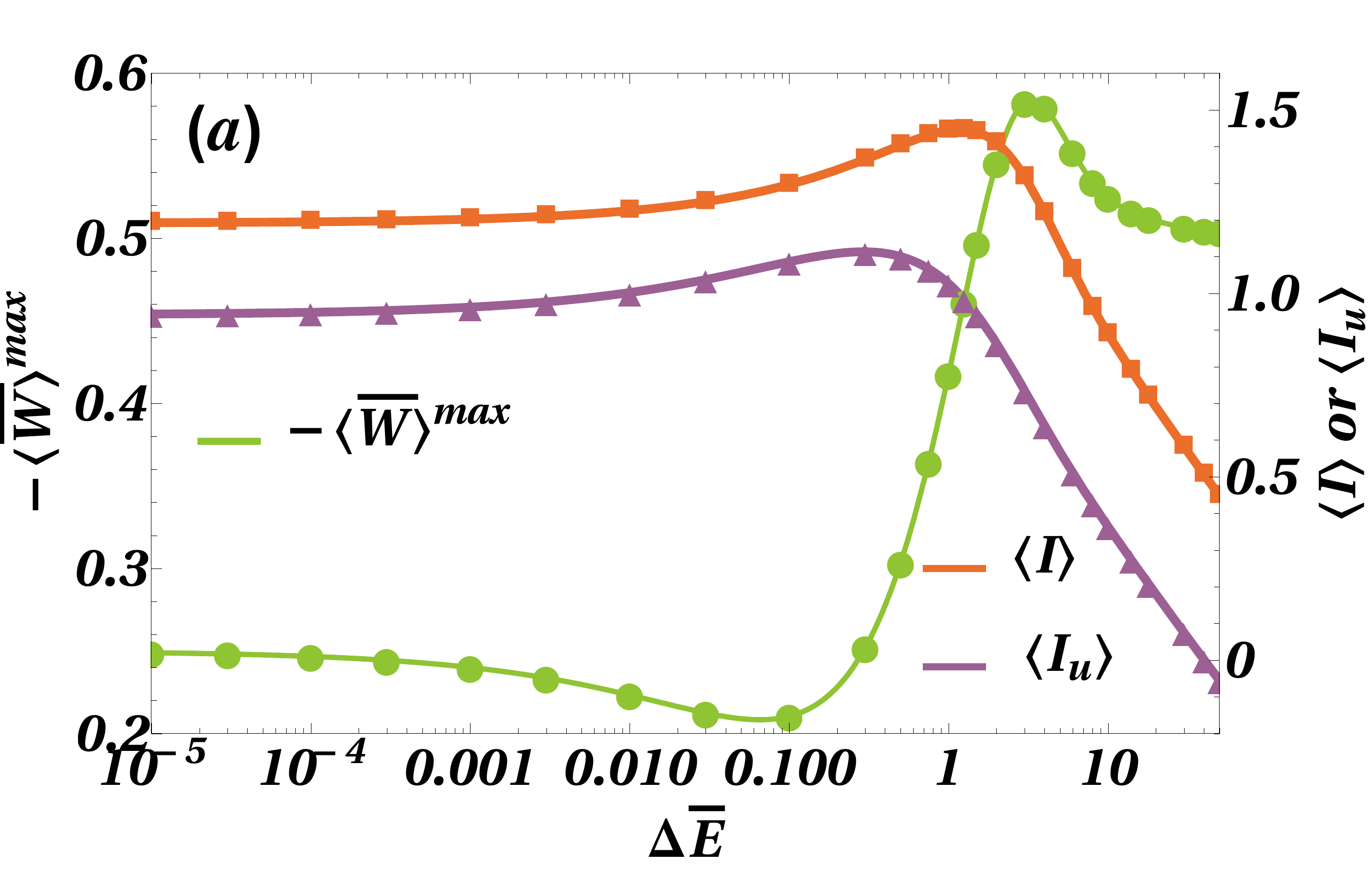} \\
      \includegraphics[width=0.42\textwidth]{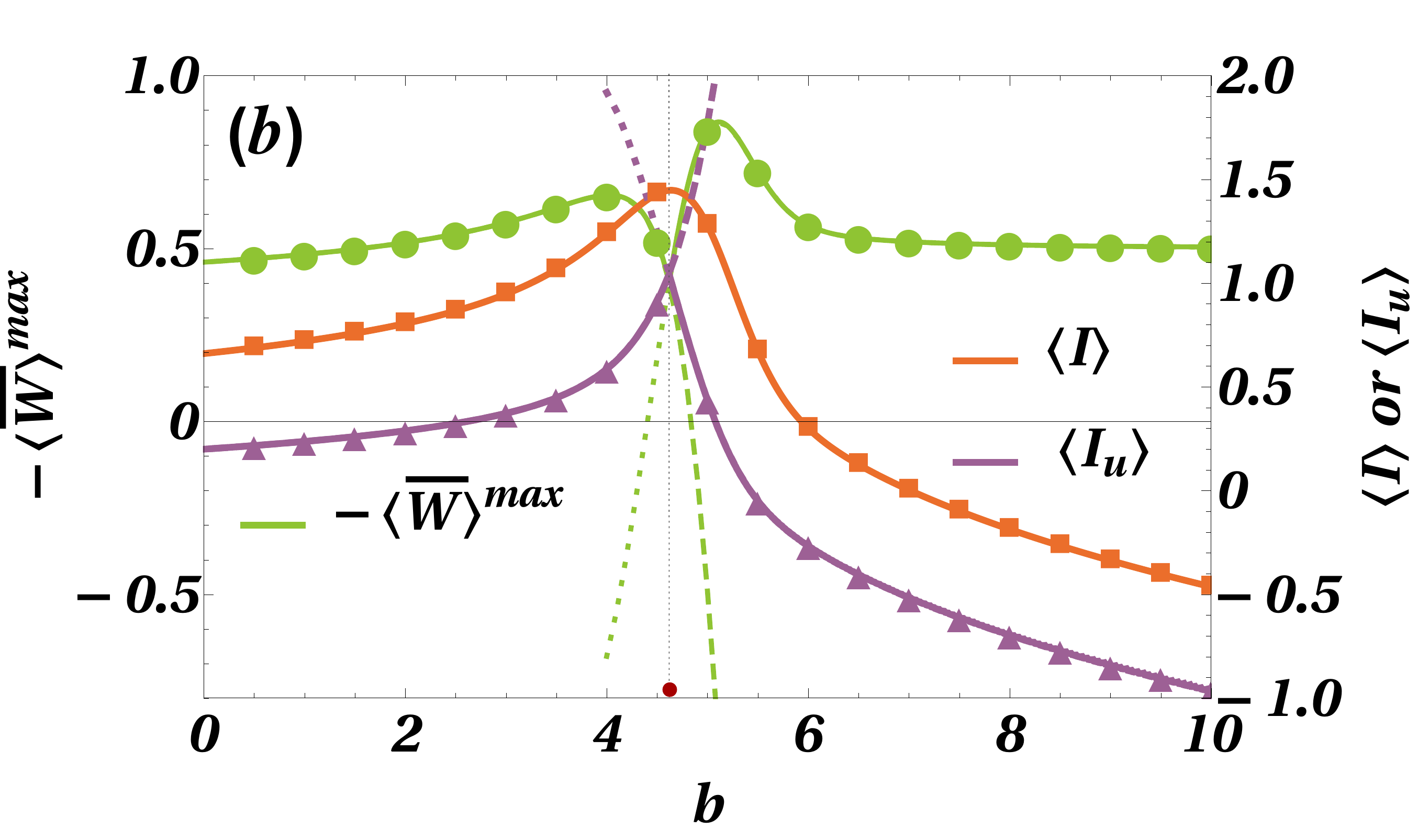} \\
     \caption{The variation upper bound of work extracted $(-\langle \overline{W} \rangle^{max})$, information $(\langle I \rangle)$ and unavailable information $(\langle I_u \rangle)$ as function of (a) scaled energy barrier $(\Delta \overline{E})$, for a BIE with bistable confinement with $b=1$, (b) quartic contribution $(b)$, for BIE with tri-stable condiment with $a=4$ and $c=1$.}
    \label{f5}
\end{figure}

Fig.~\ref{f3}(a) reveals another counterintuitive observation that the increasing bistability can raise the upper bound of output work under an optimal feedback strategy. To asses this issue further, we estimate the upper bound of work under best feedback strategy, that can be read as:
\begin{equation}\label{10}
    \begin{aligned}
       -\langle \overline {W} \rangle^{max} = 2 \epsilon - &\frac{2(4 \epsilon + 1) I_{\frac{1}{4}}\left(\epsilon \right) + 8 \epsilon \left(I_{\frac{3}{4}}\left( \epsilon \right)+I_{\frac{5}{4}}\left(\epsilon \right)\right) }{8 \left(I_{\frac{1}{4}}\left(\epsilon\right)+I_{-\frac{1}{4}}\left(\epsilon\right)\right)} \\ 
       &+ \frac{2 (4 \epsilon-1) I_{-\frac{1}{4}}\left(\epsilon \right)}{8 \left(I_{\frac{1}{4}}\left(\epsilon\right)+I_{-\frac{1}{4}}\left(\epsilon\right)\right)}.
    \end{aligned}
\end{equation}
The consequences are demonstrated in Fig.~\ref{f5}(a). We notice that the upper bound of achievable work ($-\langle \overline{W} \rangle^{\text{max}}$) changes non-trivially upon an increase in the extent of the energy barrier ($\Delta \overline{E}$) and exhibits an interesting pattern of double turn-over. The figure also shows the dependence of  $(\langle I \rangle)$ and $(\langle I_u \rangle)$ on the energy barrier under the best feedback strategy that visibly complements the variation of $-\langle \overline{W}\rangle^{max}$. In the limit of $\Delta\overline {E} \to 0$, the confinement is practically monostable $(V(x) \to \frac{b}{4}x^4)$, with a stable minimum at the potential centre. Thus, $x_d^* \to 0$ and the maximum extractable work is approximately $1/4 k_BT$ \cite{rafna_pot_2025}. The lowering of the $-\langle \overline{W} \rangle^{\text{max}}$ value with an initial rise in $\Delta \overline{E}$ can be explained in terms of the growing roughness \cite{zwanzig1988} around the potential minima \cite{rafna_pot_2025}. In the opposite limit of $\Delta\overline {E} \to high$, the particles trapped in a profound bistable potential can safely be assumed to be distributed equally in two well-defined harmonic oscillators, centred at $\pm x_min$. Therefore, the upper limit of achievable work is expected to saturate at $1/2 k_BT$. When $\Delta \overline{E} \sim 1$, particles trapped in two harmonic (almost) experiences the effect of concave curvature \cite{rafna_pot_2025, rafna_pot_2_2025} at the barrier top significantly during the relaxation process, which reduces the loss of information (lower $I_u$). Consequently,   the upper bound rises beyond the average potential of a harmonic confinement, $-\langle \overline{W} \rangle^{\text{max}} >1/2 k_BT$. Overall, two turnover, one maximum (at $\Delta \overline{E} \sim 3.3$) and one minimum  ( at $\Delta \overline{E} \sim 0.65$), is been realized (Fig.~\ref{f5} (a)). Therefore, with an appropriate depth of the potential wells and implementation of the best feedback strategy, a more efficient energy harnessing is possible, where the upper bound crosses the threshold of average potential energy.

\subsection{ Confinement with multiple hills and valleys: Feedback strategy to tune the heating-refrigeration re-entrance}

Finally, we examine the best feedback strategy for the information-energy interconversion in a potential with multiple peaks and valleys. We consider a triple well potential profile: $V(x) = \frac{a}{2}x^2-\frac{b}{4}x^4 +\frac{c}{6}x^6$, where $a$, $b$ and $c$ are  confinement constants (positive). The quartic contribution $(b)$ induces tri-stability (Fig.~\ref{fa1}(c)), as the hexic potential transitions from a single well ($b=0$) to a profound triple well ($b > \sqrt{4ac}$). 
The associated equilibrium probability distribution $(P_{eq}(x))$ changes from a unimodal to a trimodal distribution, which is shown in Fig.~\ref{fa2}(c). 
The numerical integration of Eq.~\ref{4} helps to determine the extractable work from the feedback protocol. Fig.~\ref{f3}(b) shows that the variation of $-\langle \overline{W} \rangle$ with an additional feedback distance $(x_d)$ for different shapes of the hexic confinement. In the limit of $b \ll \sqrt {4ac}$, the confinement has a rough monostable surface, centered at the origin, and $-\langle \overline{W} \rangle$ decreases monotonically with increasing $x_d$. In contrast, when $b \gg \sqrt {4ac}$, the
tri-stability is profound and $-\langle \overline{W} \rangle$ shows a nontrivial non-monotonic variation with multiple turnover with varying $x_d$.

 As expressed in Eq.~\ref{4}, the upper bound of achievable work can be found for the protocol with $x_d$ set at the position of the global potential minimum $(x_d^*=x_{min}^G)$, i.e. $x_f^*=x+x_{min}^G$. The global minimum for the chosen confinement $x_{min} ^G$ varies with the quartic contribution, for $b < 4\sqrt {ac/3}$ shows $x_{min} ^G=0$, otherwise $x_{min} ^G \neq 0$. Fig.~\ref{f3}(b) also shows that a continuous variation of $x_d$ may lead to multiple distinct heater-to-refrigerator (or vice versa) transitions, resulting in multiple re-entrance phenomena. To investigate it further, we draw a phase diagram of work extraction from feedback with an arbitrarily given triple-well potential, see Fig.~\ref{f4}(b). The phase diagram depicts that the triple-well potential, with three wells of comparable stability, the engine functions as a heater in three distinct regions of the feedback site, all close to the local potential minima, which results in multiple heater-to-refrigeration re-entrance phenomena. One can estimate the associated variation in $\langle I \rangle$ and $\langle I_u \rangle$ (Fig.~\ref{fb1} (c)) to confirm that the information lost $(\langle I_u \rangle)$ during the relaxation under different $x_d$, controls the relative dominance of $\langle I \rangle$ and $ \langle I_u \rangle$, and their crossing. For confinements with $ b \ge 4\sqrt {ac/3}$, the $\langle I_u \rangle$ shows a non-monotonic variation with rising $x_d$ characterised by multiple turn-over behaviour. The minimum information loss consistently occurs when the feedback distance matches the coordinate of the global potential minimum $(x_d^*=x_{min}^G)$. 

Finally, to comprehend the influence of the best feedback strategy on the upper bound of work extraction, we calculate the variation of $-\langle \overline{W} \rangle^{max}$ as a function of quartic contribution $(b)$ of the triple well potential (Fig.~\ref{f5}(b)). In the limit of $b\rightarrow 0$, the confinement is essentially a monostable harmonic potential with a minor contribution of $\sim x^6$ term  (as we have taken $a >> c$), thus $-\langle \overline{W} \rangle^{max}$ can be obtained as little less than $1/2$. In contrast, in the limit of $b\rightarrow high$, particles are confined in the close vicinity of the minima of three wells, experiencing an environment of harmonic potential near each deep, resulting in $-\langle \overline{W} \rangle^{max} \sim 1/2$. The discontinuity in variation arises due to the change in global minima (which equals the optimal additional feedback distance), i.e. $x_f^*=x$ for $b < 4 \sqrt{ac/3}$, otherwise $x_f*=x + x_{min}$. The variation also affirms the very fact that the contribution of $\langle V(x) \rangle$ and  $V(x_d)$ in determining the upper bound of output work may vary non-trivially with a change in $b$ that enables higher work extraction than a harmonic trap with a triple wells of moderate depth, due to the comparatively lower information loss during relaxation. 
\section{Summary}
In conclusion, we have generalised a feedback strategy for the most efficient information-energy exchange in a BIE operating in an arbitrary confined potential. 
We find that the best work extraction protocol involves shifting the potential center to the feedback site $x_f^* = x + x_{\min}^G$, where $x$ is the measurement outcome and $x_{\min}^G$ is the global potential minimum position (assuming $V(x_d=x_{min}^G) < 0$). The relative dominance of the average potential energy$(\langle V(x) \rangle)$ and the potential with an additional feedback distance ($(V(x_d)\; \text{with} \;x_d=x-x_f$) determines the parameter zone where the BIE acts as an engine. The process is heating only if $(V(x_d) < \langle V(x) \rangle)$, otherwise a refrigerator. Depending on the complexity of the potential surface, one may have multiple occurrences of engine-refrigerator reentrance (multiple  $x_d^{inv}$). We then explain the consequences of the best feedback strategy, using three geometrically different confinements as illustrative examples.
The BIE under single-well confinement with varying concavity $(V(x) = a|x|^n)$, the symmetric protocol serves as optimal strategy yielding the upper limit of work extraction, i.e. for $x_f=x\; (x_d=0)$, $- \langle \overline{W} \rangle^{max} = \frac{1}{n}$, which is equal to the average potential energy.
We find that confinement with a perturbed centre (a bistable one) or a triple-well potential may exhibit multiple good feedback conditions and open the possibility of refrigerator-to-heater transitions and their reentrance. Once the shape of the double or triple-well potential surface is chosen carefully,  the upper bound of the achievable work obtained from BIE may surpass the average potential energy of the system.
\begin{acknowledgments}
RR acknowledges IIT Tirupati and DST-INSPIRE for fellowship (Project No.DST/INSPIRE/03/2021/002138). DM thanks SERB (Project No.  ECR/2018/002830/CS), Department of Science and Technology, Government of India, for financial support. 
\end{acknowledgments}
\section*{Data Availability}
The data that support the findings of this study are available within the article.

\bibliographystyle{achemso}
\bibliography{References}

\appendix
\counterwithin{figure}{section}

\section{Analysis of probability distribution and Information processing}
\label{appA}
\begin{figure*}[!htb]
\begin{minipage}[b]{0.5\linewidth}
\centering
\includegraphics[width=1\textwidth]{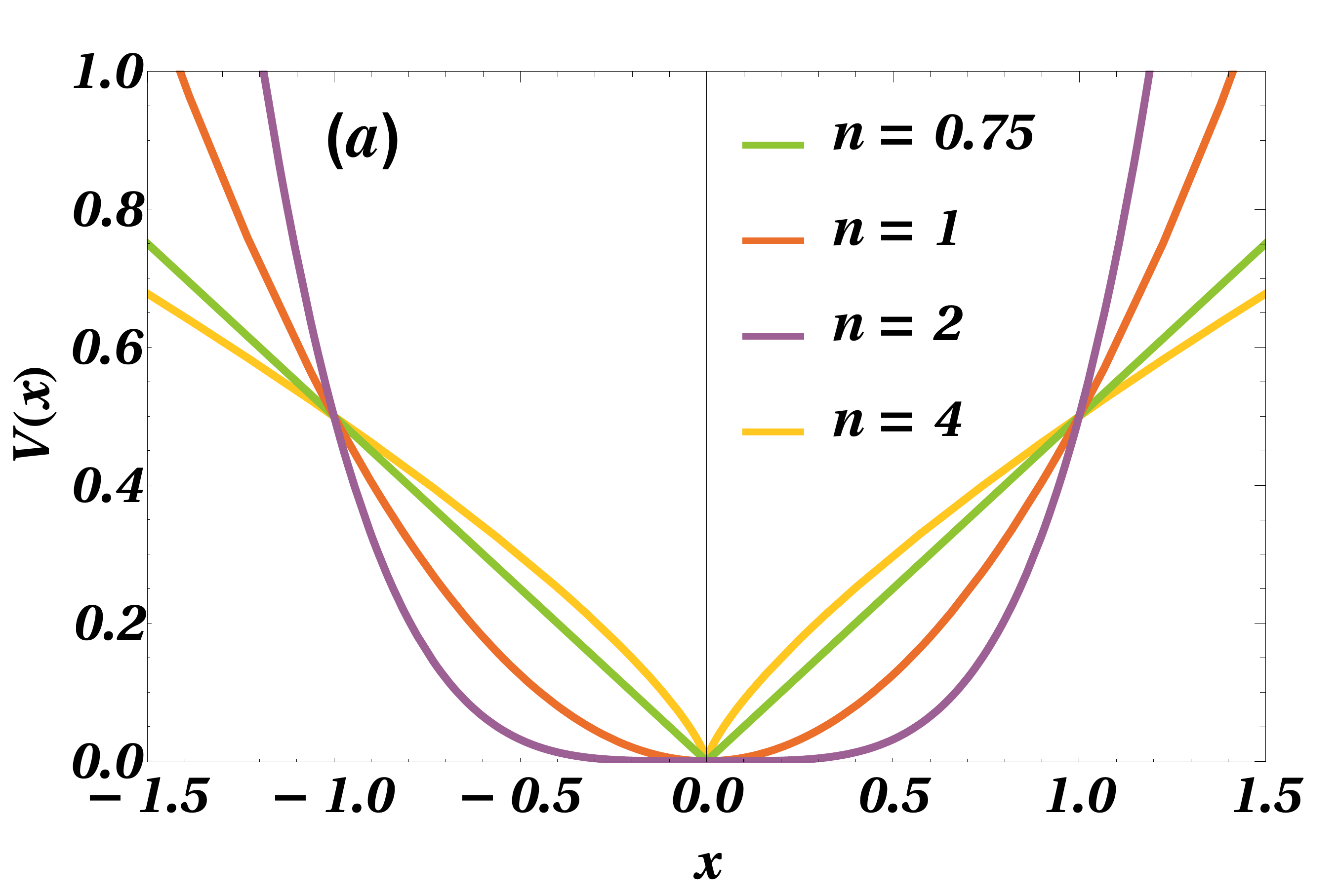}
\end{minipage}
\hspace{0.01cm}
\begin{minipage}[b]{0.5\linewidth}
\centering
\includegraphics[width=1\textwidth]{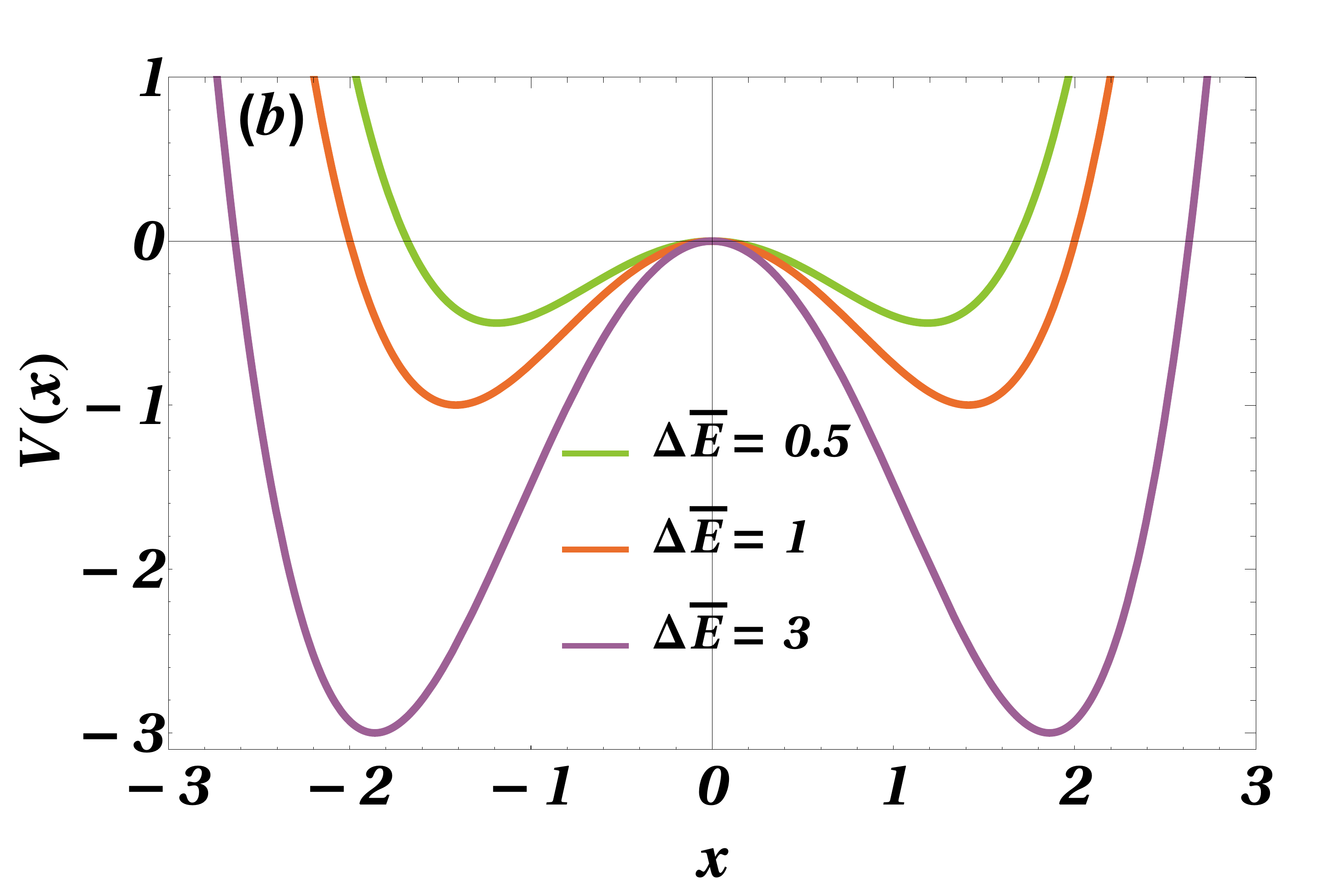}
\end{minipage}
\hspace{0.01cm}
\begin{minipage}[b]{0.5\linewidth}
\centering
\includegraphics[width=1\textwidth]{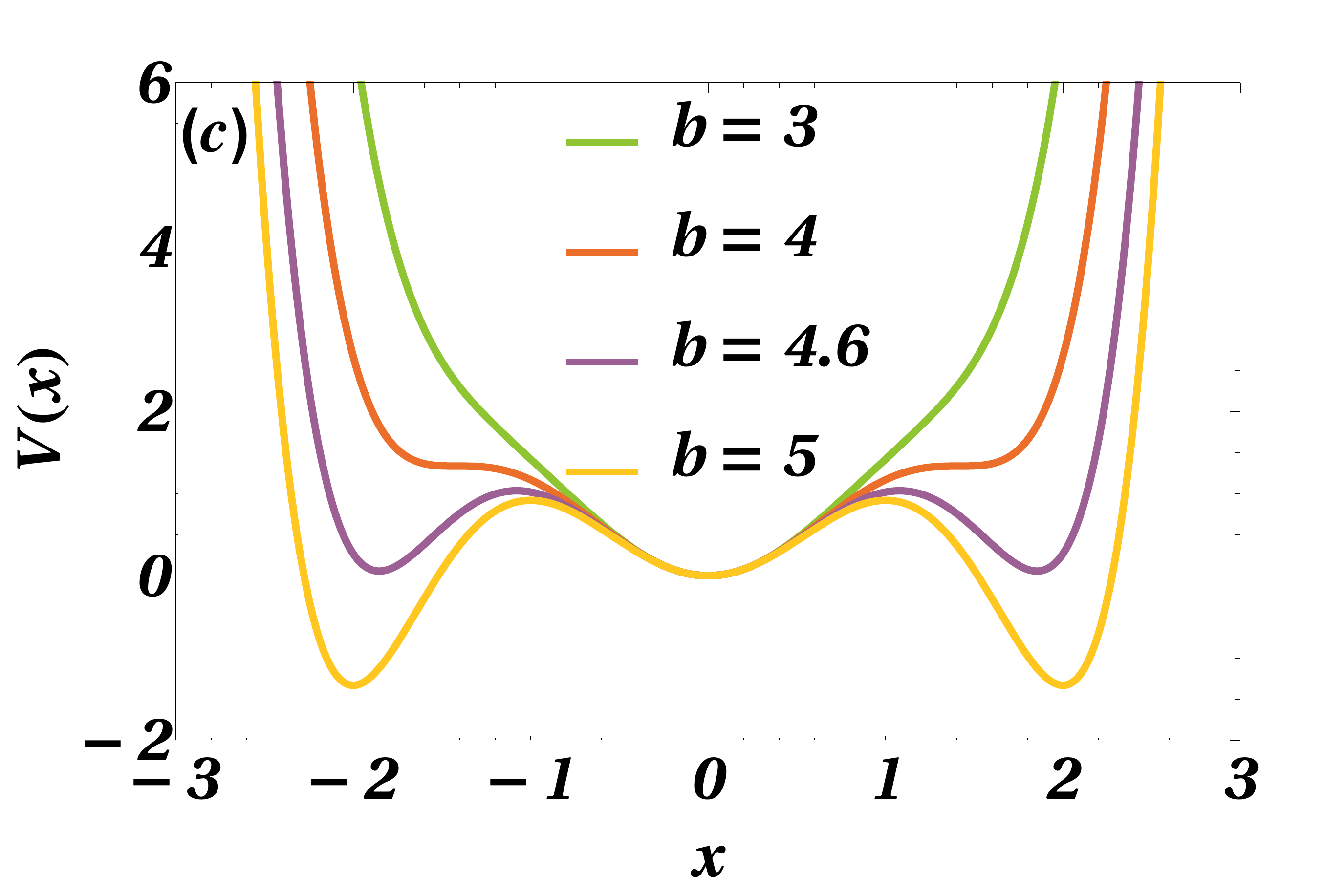}
\end{minipage}
\caption{ (a) Monostable potential $V(x)=a|x|^n$ as a function of $x$ with different values of exponent ($n$) and with $a=0.5$. (b) Bistable trapping $V(x)= -\frac{a}{2}x^2+\frac{b}{4}x^4$ as a function of $x$ under varying the scaled barrier height $\Delta \overline{E}$ ($\Delta E = a^2/4b$), and with $b=1$. (c)Triple well potential $V(x) = \frac{a}{2}x^2-\frac{b}{4}x^4 +\frac{c}{6}x^6$ as a function of $x$ with varying quartic contribution $(b)$, with $a=4$ and $c=1$.}
\label{fa1}
\end{figure*}

\begin{figure*}[!htb]
\begin{minipage}[b]{0.5\linewidth}
\centering
\includegraphics[width=1\textwidth]{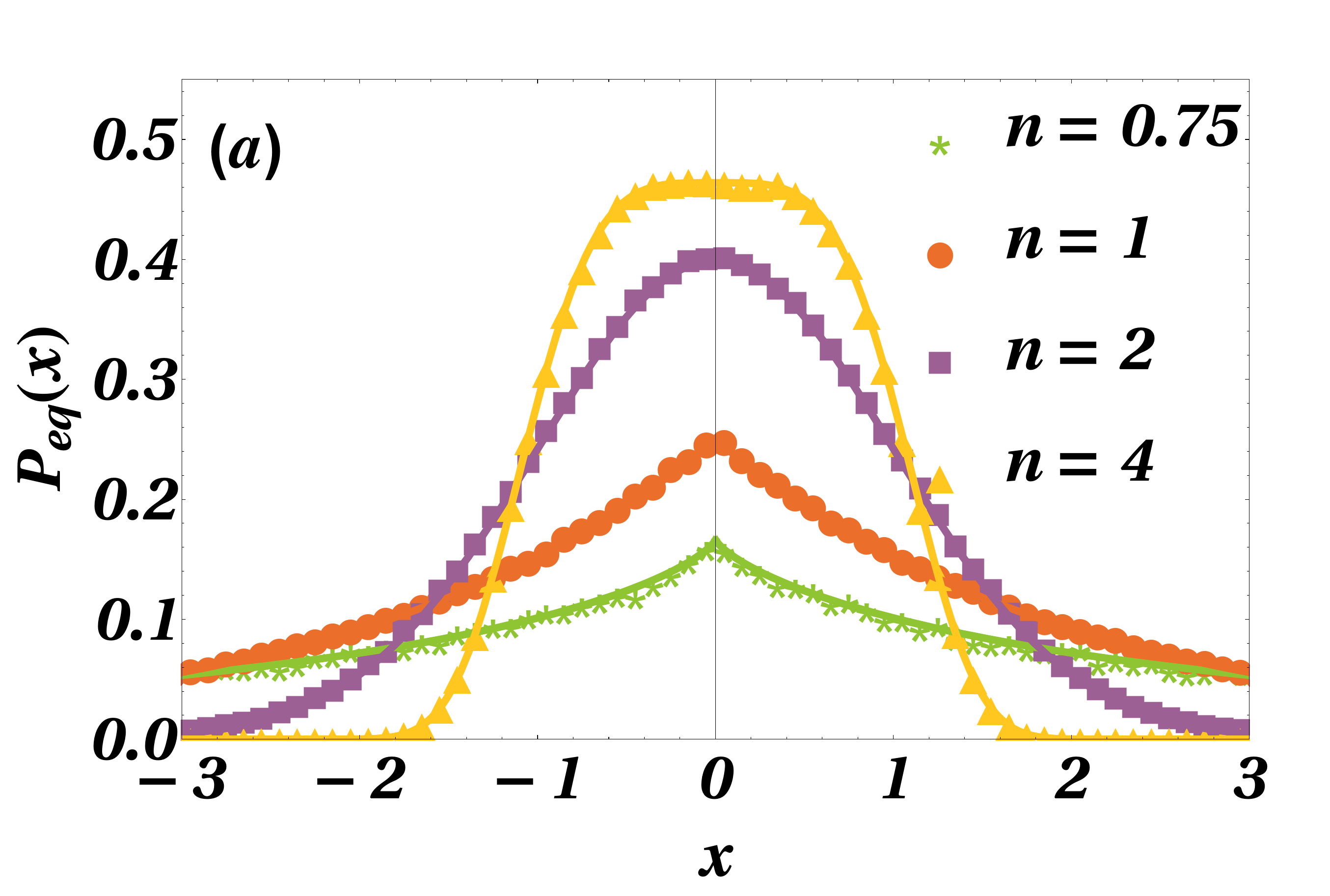}
\end{minipage}
\hspace{0.01cm}
\begin{minipage}[b]{0.5\linewidth}
\centering
\includegraphics[width=1\textwidth]{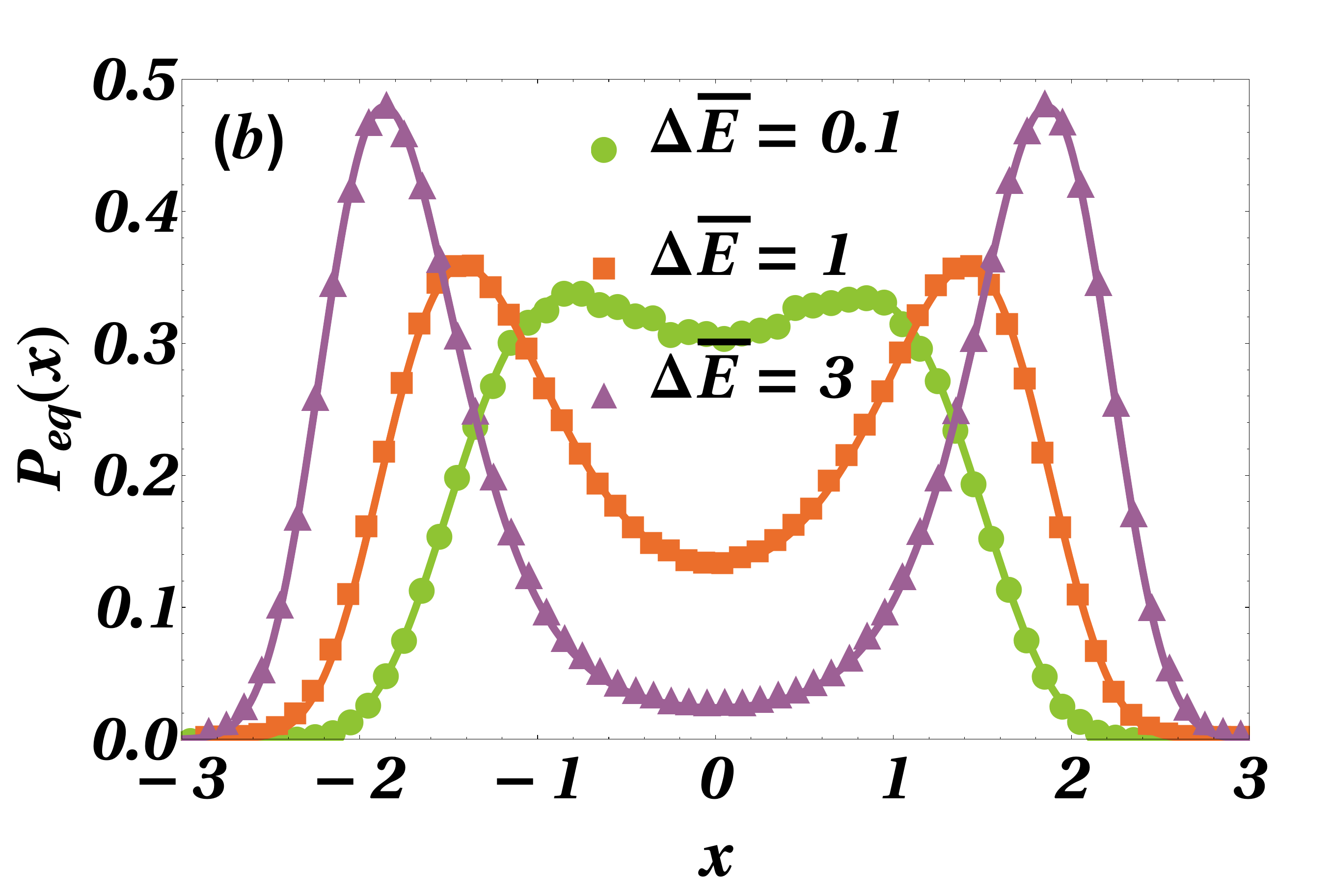}
\end{minipage}
\hspace{0.01cm}
\begin{minipage}[b]{0.5\linewidth}
\centering
\includegraphics[width=1\textwidth]{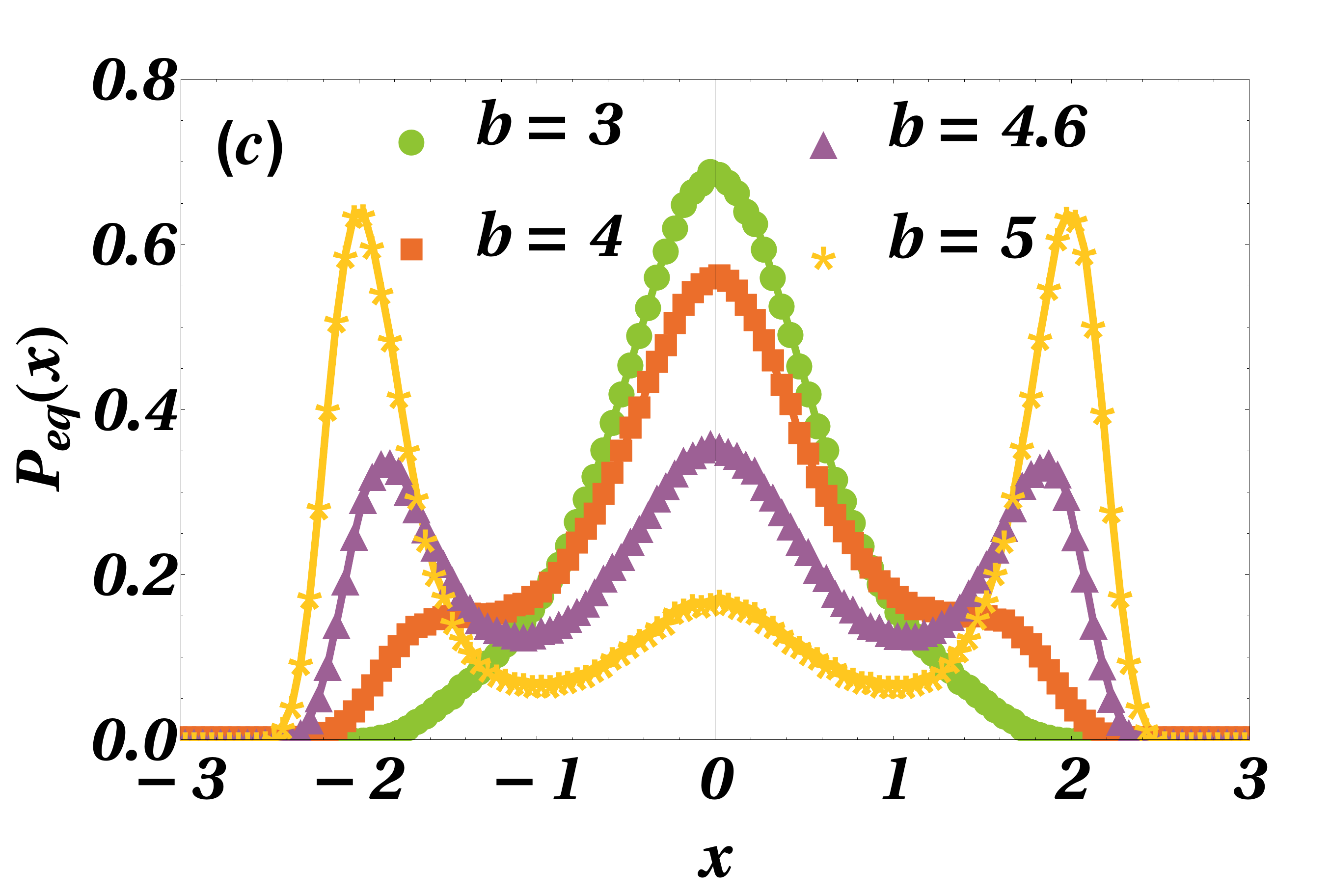}
\end{minipage}
\caption{The equilibrium probability distribution $(P_{eq}(x))$ of (a) monostable potential as a function of $x$ with different values of exponent ($n$). (b) bistable trapping as a function of $x$ under varying the scaled barrier height $(\Delta \overline{E})$ (c) tri-stable potential, as a function of $x$ with varying quartic contribution $(b)$. The potential parameters are kept the same as Fig.~\ref{fa1}. In all cases, solid lines represent the theoretical predictions and points are obtained from numerical simulation (Eq.~\ref{1}). The parameter set chosen: $\gamma = 1$ and $k_BT=1$.}
\label{fa2}
\end{figure*}

For a Brownian particle confined within the monostable potential of the form $V(x) =a |x|^n$ (as shown in Fig.~\ref{fa1}(a)), one can  derive the equilibrium probability distribution of the particle’s position following Eq.~\ref{2} as:
\begin{equation}\label{A1}
    \begin{aligned}
      P_{eq}(x)= \mathcal{N}_m \exp  [- \overline{a}|x|^n  ], \; \text{with} \;
 \mathcal{N}_m =   \frac{\overline{a}^{\frac{1}{n}}}{2 \Gamma (1+\frac{1}{n})}.
    \end{aligned}
\end{equation}
Here, $\Gamma(z)$ represents the gamma function, defined by the integral $\Gamma(z) = \int_0^\infty t^{z-1} e^{-t}  dt$. 
As depicted in Fig.~\ref{fa2}(a), $P_{eq}(x)$ has a symmetric unimodal distribution. The shape of the peak and tail behaviour changes with the extent of concavity of the confining potential.

For potential in quartic form: $V(x)= -\frac{a}{2}x^2 + \frac{b}{4} x^4$ exhibit two global minima at $x = \pm x_{min}$ $(x_{min}= \sqrt{\frac{a}{b}})$ and one global maxima at $x_{max}=0$. The equilibrium probability distribution of the particle's position under the influence of this external bistable potential can be derived from Eq.~\ref{2} as:
\begin{equation}\label{A2}
    \begin{aligned}
      P_{eq}(x) &= \mathcal{N}_b  \exp \bigg [ \frac{\overline{a}}{2}x^2 - \frac{\overline{b}}{4} x^4 \bigg ], \\
      \mathcal{N}_b 
 &= \frac{e^{-\epsilon}}{\pi} \bigg ( \frac{2\overline{b}}{\epsilon} \bigg )^{\frac{1}{4}} \big [ I_{\frac{1}{4}} (\epsilon) +I_{-\frac{1}{4}} (\epsilon) \big ],
    \end{aligned}
\end{equation}
with the notation bearing the meaning as: $\epsilon = \Delta \overline{E}/2$, $\Delta \overline{E} = \Delta E / k_BT$ and $I_{\nu}(z)$ is the modified Bessel function of the first kind of the form, $I_{\nu}(z) = \sum_{k=0}^{\infty} \frac{(z/2)^{2k+\nu}}{\Gamma [k+ \nu + 1] k!}$. The Fig.~\ref{fa2}(b) shows the alteration in $P_{eq}(x)$ corresponding to the alteration in the scaled energy barrier of the quartic potential. In the same lane, Fig.~\ref{fa2}(c) depicts the variation of $P_{eq}(x)$ as a function of $x$ for different values of potential parameter $b$ for BIE with triple well confinement.

The potential energy of hexic form: $V(x) = \frac{a}{2}x^2-\frac{b}{4}x^4 +\frac{c}{6}x^6$ displays tri-stability when the condition $b> \sqrt{4ac}$ is satisfied. Under this condition, the potential has two maxima symmetrically located at $x = \pm x_{max}$ and three minima located at $x=0$ and $ x = \pm x_{min}$, where:
\begin{equation}\label{A3}
\begin{aligned}
     x_{max}=\sqrt{\frac{b-\sqrt{b^2-4 a c}}{2 c}}; \\
    x_{min}=\sqrt{\frac{b+\sqrt{b^2-4 a c}}{2 c}}.
\end{aligned}
\end{equation}
 Interestingly, for $b < 4 \sqrt{ac/3}$, the global minima can be found at the origin $x_{min}^G=0$.  However, $x_{min}^G=\pm x_{min}$ or $b < 4 \sqrt{ac/3}$. We obtain $P_{eq}(x)$ using numerical integration. 
 \begin{equation}\label{A3a}
      -\langle W \rangle = -\int_{-\infty}^{\infty} dx P_{eq}(x)\left(V(x)-V(x_d)\right).
\end{equation}

For BIE with monostable confinement the average acquired information $(\langle I \rangle)$ and unavailable information $(\langle I_u \rangle)$ can be derived using Eq.~\ref{5}, \ref{6} and \ref{A1} as:
\begin{equation}\label{B1}
    \begin{aligned}
        \langle I \rangle = -\ln [\mathcal{N}_m ] + \frac{1}{n}, \\
        \langle I_u \rangle = \overline{a} |x_d|^n -\ln [\mathcal{N}_m ].
    \end{aligned}
\end{equation}\\

Following Eq.~\ref{5}, \ref{6} and \ref{A2}, the average acquired information $(\langle I \rangle)$ and unavailable information $(\langle I_u \rangle)$ for the protocol with the bi-stable confinement will be of the form: 
\begin{equation}\label{B2}
    \begin{aligned}
\langle I \rangle &= -\ln  [ \mathcal{N}_b] + \frac{[4\epsilon - 1] I_{-\frac{1}{4}} \left( \epsilon \right)
-[4 \epsilon +1] I_{\frac{1}{4}}\left(\epsilon\right)}
{4  [ I_{\frac{1}{4}}\left( \epsilon \right ) +I_{-\frac{1}{4}}\left( \epsilon \right ) ]}
\\ & \quad -\frac{ \epsilon [ I_{\frac{3}{4}}(\epsilon )+I_{\frac{5}{4}}\ (\epsilon ) ]}{I_{\frac{1}{4}}(\epsilon )+I_{-\frac{1}{4}} (\epsilon)}, \\
\langle I_u \rangle &= - \sqrt{2 \overline{b} \epsilon} x_d^2 + \frac{\overline{b}}{4} x_{d}^4  -\ln [ \mathcal{N}_b].
    \end{aligned}
\end{equation}
Similarly to the calculation of the exciting work, for a triple-well potential confinement, we analytically estimate the average acquired information $(\langle I \rangle)$ and unavailable information $(\langle I_u \rangle)$, performing the required numerical integrations.

\section{Dimensionless description of Langevin Equation and Numerical simulation details}
In this paper, we perform the numerical estimation of the equilibrium probability distribution of particle $P_{eq}(x)$  using the overdamped Langevin equation.
The overdamped Langevin equation of motion for a particle confined in an arbitrary external potential $(V(x))$ can be written in its conventional form:
\begin{equation}\label{A4}
    \gamma \dot{x} = -V^{'}(x) + \sqrt{2\gamma k_BT}  \eta(t).
\end{equation}
The particle position is denoted by $x$, time by $t$, $\gamma$ represents the friction coefficient, and $T$ is the temperature (in absolute scale) of the thermal reservoir and $k_B$ is the Boltzmann constant; thermal noise is modeled as zero-mean Gaussian white noise $\zeta(t)$. $\zeta(t)$ has a dimension of $t^{-1/2}$.
We consider a dimensionless description of the process by scaling all the variables and parameters by suitable reference scaling factors as reported in recent experimental studies \cite{Paneru2018prl, Paneru2020natcommun}. We consider a reference energy scale $E_r$ is defined as $k_B T_R$, with $T_R = 293\,\text{K}$. The reference friction coefficient is set to $\gamma_r = 18.8\,\text{nN}.\text{m}^{-1}.\text{s}$. The position of the particle may be scaled by a characteristic length scale $x_r \approx 20\,\text{nm}$, and the time by $t_r \approx 2\,\text{ms}$. This consideration guides us to scale the potential parameters. For example, the reference stiffness scale for a harmonic potential becomes $E_r / x_r^2$, equivalent to $10\,\text{pn}. \mu \text{m}^{-1}$. Now we can express energy, friction coefficient, length and time in dimensionless forms as $\tilde{E} = E/E_r$, $\tilde{\gamma} = \gamma/\gamma_r$, $\tilde{x} = x/x_r$, and $\tilde{t} = t/t_r$, respectively. In the main text (Eq. 1), we omit the tilde sign for simplicity.

The particle trajectories are generated using the modified Euler method \cite{hildebrand1987} with a time increment of $10^{-3}$ units. To model Gaussian white noise, the Box-Muller algorithm is employed \cite{box1958}. The ensemble averages are computed over approximately $10^7$ trajectories. When analytical solutions are not possible, we use Simpson’s one-third rule for numerical integration \cite{hildebrand1987}. We kept the $\gamma$ and $ k_B T$ values as unity (unaltered) throughout the manuscript. Details of other parameters can be found in the figure captions. 

\begin{figure*}[!htb]
\begin{minipage}[b]{0.5\linewidth}
\centering
\includegraphics[width=1\textwidth]{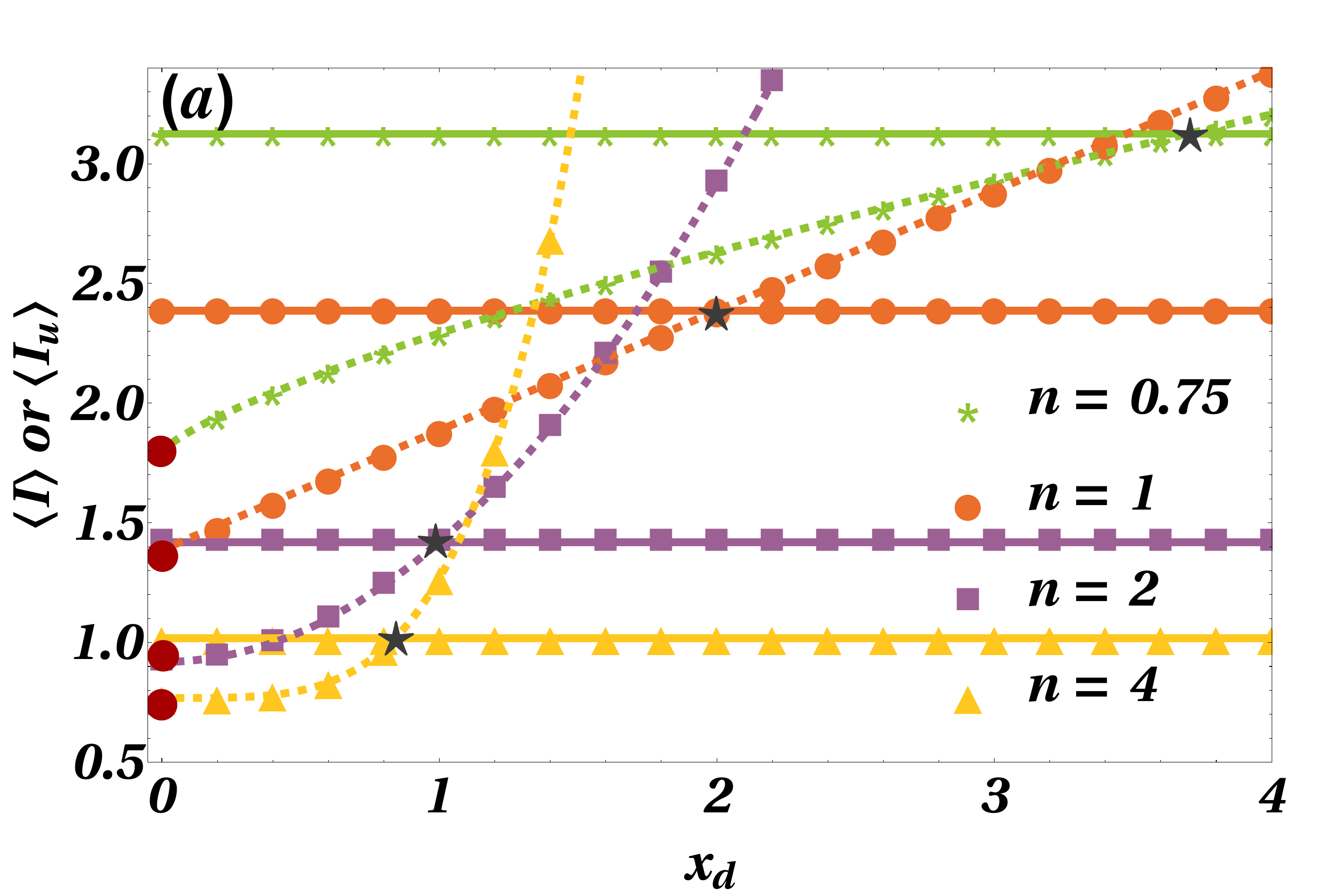}
\end{minipage}
\hspace{0.01cm}
\begin{minipage}[b]{0.5\linewidth}
\centering
\includegraphics[width=1\textwidth]{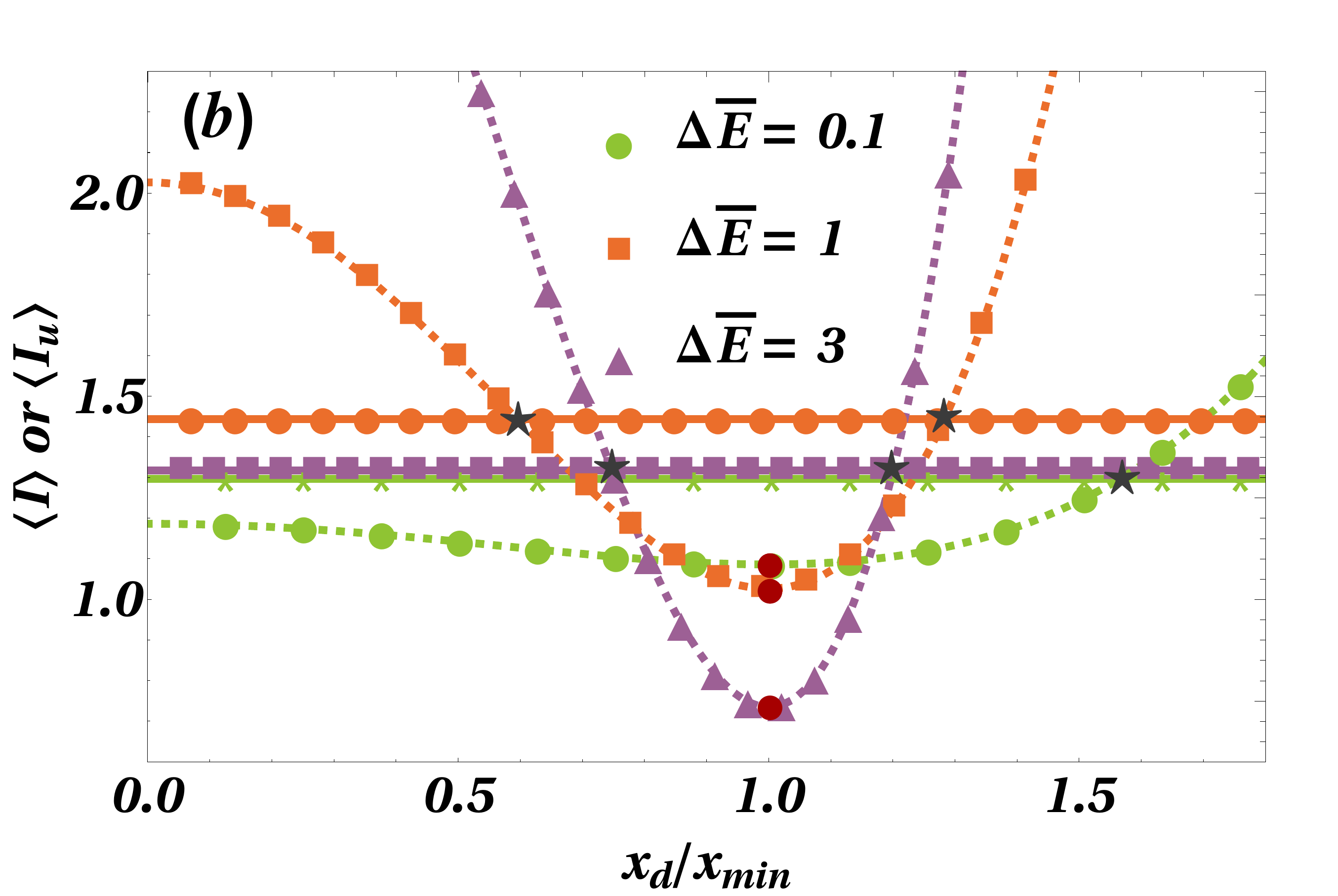}
\end{minipage}
\hspace{0.01cm}
\begin{minipage}[b]{0.5\linewidth}
\centering
\includegraphics[width=1\textwidth]{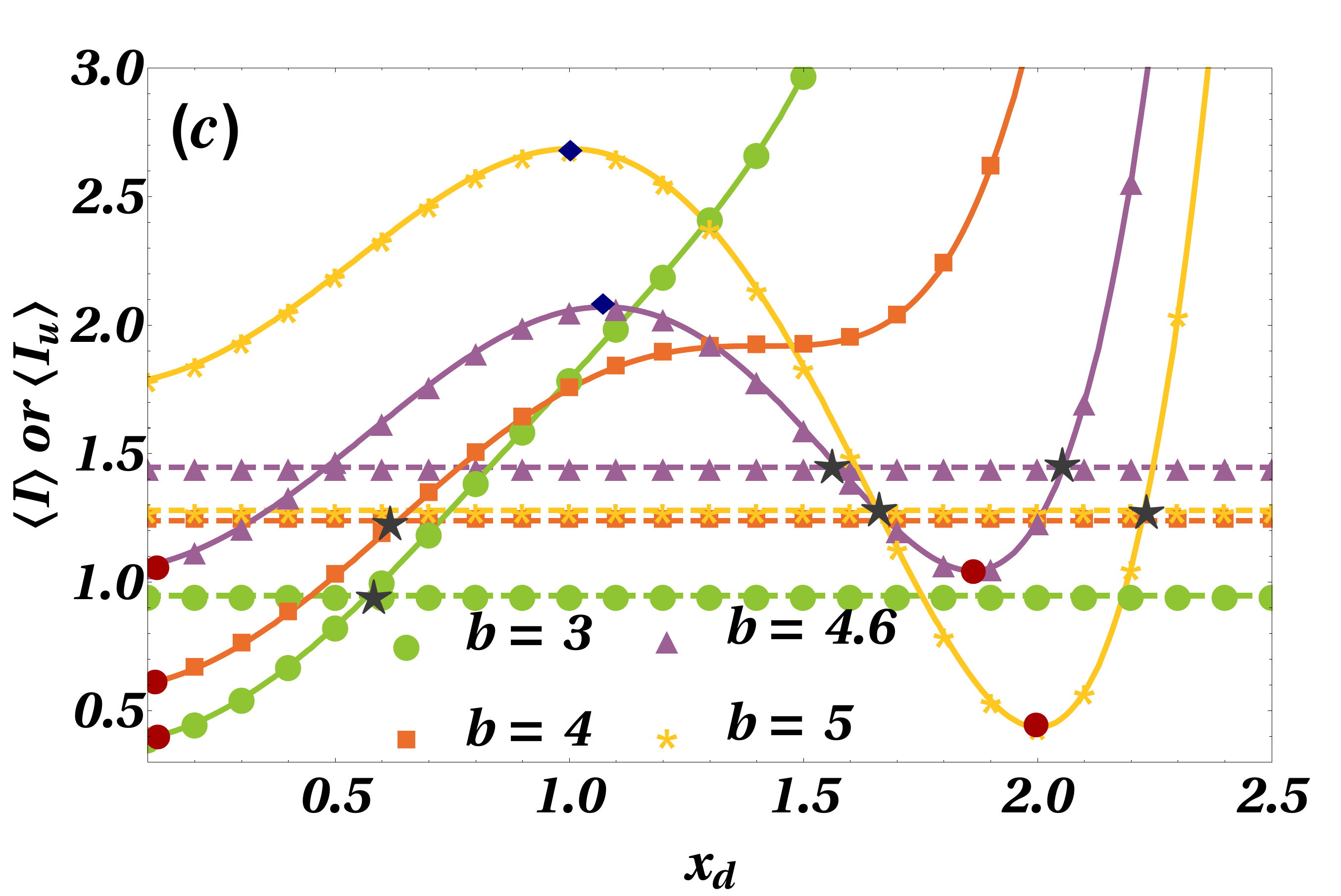}
\end{minipage}
\caption{Variation of average acquired information $(\langle I \rangle \text{,\; filled shapes-solid lines })$ and average unavailable information $(\langle I_u \rangle  \text{, \;filled shapes-dotted lines})$ as a function of  (a) additional feedback distance $(x_d)$, for different values of power exponent $n$. (b) scaled additional distance $(x_d/x_{min})$ for potential with varying $\Delta \overline{E}$. (c) as a function of additional distance $(x_d)$ for potential with varying quartic contribution $(b)$. 
 The potential parameters are kept the same as Fig.~\ref{fa1}. Red-colored circles indicate additional distance $(x_d^*)$ criteria to obtain output maxima $(-\langle \overline{W} \rangle^{max})$, dark grey-colored stars represent the output inversion points $(x_d^{inv})$ and blue-colored diamonds indicate feedback conditions to obtain the minima of work output. For all cases, the parameter set is chosen: $\gamma =1$ and $k_BT=1$. }
\label{fb1}
\end{figure*}


\end{document}